\documentclass[useAMS,usenatbib]{mnras}

\usepackage{amsmath}
\usepackage{amsfonts}
\usepackage{amssymb}
\usepackage{graphicx}

\usepackage[normalem]{ulem}
\usepackage{RefJ}

\usepackage{empheq}

\usepackage{enumitem}

\usepackage{hyperref}
\usepackage[]{natbib}

\usepackage{widetext}

\newcommand{\dv}{\vec{\nabla} \cdot}

\newcommand{\dd}{\mathrm{d}}

\newcommand{\m}{\scriptscriptstyle -}
\newcommand{\p}{\scriptscriptstyle +}

\let\vec\boldsymbol 

\usepackage[dvipsnames]{color}

\usepackage{array}
\newcolumntype{T}{>{\tiny}l} 
\newcolumntype{H}{>{\Huge}l} 

\title[An MHD spectral theory approach to Jeans' magnetized gravitational instability]{An MHD spectral theory approach to Jeans' magnetized gravitational instability}
\author[J.-B. Durrive, R. Keppens, and M. Langer]{Jean-Baptiste Durrive$^{1,2,3}$\thanks{E-mail:jdurrive@protonmail.com}, Rony Keppens$^{3}$, and Mathieu Langer$^{4}$\\
$^{1}$ Institut de recherche en astrophysique et plan\'etologie - Universit\'e Toulouse III - Paul Sabatier, Observatoire Midi-Pyr\'en\'ees,\\
Centre National de la Recherche Scientifique, UMR5277 - France\\
$^2$ Laboratoire de Physique de l'Ecole normale sup\'erieure, ENS, Universit\'e PSL, CNRS, Sorbonne Universit\'e, Universit\'e de Paris,\\ F-75005 Paris, France\\
$^3$ Centre for mathematical Plasma-Astrophysics, Celestijnenlaan 200B, 3001 Leuven, KU Leuven, Belgium\\
$^4$ Universit\'e Paris-Saclay, CNRS, Institut d'Astrophysique Spatiale, B\^atiment 121, 91405 Orsay, France\\}

\begin{document}

\date{Accepted 2021 June 7. Received 2021 June 4; in original form 2021 April 9}

\pagerange{\pageref{firstpage}--\pageref{lastpage}} \pubyear{2020}

\maketitle

\label{firstpage}

\begin{abstract}
In this paper, we revisit the governing equations for linear magnetohydrodynamic (MHD) waves and instabilities existing within a magnetized, plane-parallel, self-gravitating slab. Our approach allows for fully non-uniformly magnetized slabs, which deviate from isothermal conditions, such that the well-known Alfv\'en and slow continuous spectra enter the description.  We generalize modern MHD textbook treatments, by showing how self-gravity enters the MHD wave equation, beyond the frequently adopted Cowling approximation. This clarifies how Jeans' instability generalizes from hydro to magnetohydrodynamic conditions without assuming the usual Jeans' swindle approach. Our main contribution lies in reformulating the completely general governing wave equations in a number of mathematically equivalent forms, ranging from a coupled Sturm-Liouville formulation, to a Hamiltonian formulation linked to coupled harmonic oscillators, up to a convenient matrix differential form.
The latter allows us to derive analytically the eigenfunctions of a magnetized, self-gravitating thin slab. In addition, as an example we give the exact closed form dispersion relations for the hydrodynamical p- and Jeans-unstable modes, with the latter demonstrating how the Cowling approximation modifies due to a proper treatment of self-gravity. The various reformulations of the MHD wave equation open up new avenues for future MHD spectral studies of instabilities as relevant for cosmic filament formation, which can e.g. use modern formal solution strategies tailored to solve coupled Sturm-Liouville or harmonic oscillator problems.
\end{abstract}


\ \vspace{-1.7cm}

\section{Introduction}

\subsection{Motivations from Astrophysics and Cosmology}

Sheet-like and filamentary structures of matter are ubiquitous in the Universe. For example, they are routinely observed in the interstellar medium of our Galaxy, in which giant molecular clouds are shaped by the combined action of gravity, supernovae explosions, thermal instability, cloud-cloud collisions, turbulence and magnetic fields 
\cite[][]{SchneiderElmegreen79,BallyEtAl87,MizunoEtAl95,Hartmann02,Myers09,PudritzKevlahan13,Andre15}.
Similarly, at cosmological scales, gravity organizes matter into a cosmic web of voids delineated by cosmological walls and filaments, as demonstrated by numerical simulations \citep[][]{KlypinShandarin83,KlarMucket10}. At the nodes of this web lie galaxy clusters, which are supplied with matter, baryonic and dark, flowing along the filaments that interconnect them. Part of this accretion occurs intermittently \citep[][]{DekelEtAl09_a,DekelEtAl09_b,KeresEtAl09,SanchezAlmeidaEtAl14}, suggesting that denser clumps of matter might form not only within galaxy clusters, but also either in the voids, the walls or the filaments of the cosmic web.
While a fraction of these clumps may in fact be numerical artifacts, most of the clumps are believed to have a true physical origin \citep[][]{Springel10,HobbsEtAl13,NelsonEtAl13,HobbsEtAl16}.
Do baryons in cosmological walls and filaments fragment due to their own gravitational instability, or are these (numerically) observed gas clumps exclusively the product of the growth of primordial overdensities, as baryons fall into the gravitational potential of collapsed dark matter halos that they are embedded into?
The general motivation of this paper is to contribute to the study of the fragmentation of magnetized and unmagnetized self-gravitating gas structures, in order to predict the size and growth rates of formation of clumps from astrophysical to cosmological scales. Such predictions are essential to better understand star and galaxy formation.

Many different instabilities may in principle give rise to this fragmentation. The thermal, Rayleigh-Taylor or Kelvin-Helmholtz instabilities surely play a role, for example in the cosmological context, in the denser environments of massive haloes \citep[e.g.][]{KeresHernquist09}. But another well identified universal actor at play, which is the main focus of the present work, is Jeans' gravitational instability, including magnetic fields, given their important dynamical role \citep{Parker79,Cox05}.
In the literature, `gravitational instability' may refer to the convective instability or the Rayleigh-Taylor instability, but here we deliberately choose a closure relation which switches-off convection, because our focus is on Jeans' gravitational instability.

\subsection{Brief summary on Jeans' instability analysis}

The analysis of Jeans' instability in astrophysics has a very long history.
References of historical importance studying the equilibrium states of self-gravitating structures include \cite{Ledoux1951} for planar structures, and \cite{Ostriker64a} for cylinders, and more recently an extremely detailed study of polytropes has been performed by \cite{Horedt04}.
As for the stability of these equilibria, a plethora of studies could be quoted, such that the following list is by no means exhaustive.
Historically, the investigation of gravitational instability was triggered by the works of Jeans \citep[e.g.][]{Jeans28}. Later the stability of sheet-like structures has been explored by \cite{Ledoux1951} and extended by \cite{Simon1965b} numerically.
Effects of deviations from isothermality can be found for instance in \cite{GoldreichLyndenBell65}, in which stability criteria for pressure bounded, uniformly rotating polytropic sheets are obtained. Gradually, more and more ingredients relevant to describing astrophysical and cosmological environments were taken into account. The effects of an external pressure \citep[e.g.][]{ElmegreenElmegreen78,MiyamaEtAl87a,MiyamaEtAl87b,NaritaEtAl88}, of uniform and differential rotation \citep[][]{Safronov1960,Simon65a,NaritaEtAl88,PapaloizouSavonije91,BurkertHartmann04}, of flow \citep[][]{Lacey89}, of the background expansion of the Universe and the dark matter component \citep[][]{Umemura93,AnninosEtAl95,HosokawaEtAl00}, of the possible advent of convective instability \citep[e.g.][]{MamatsashviliRice10,BreysseEtal14}, of the local expansion (or collapse) of the structure \citep[][]{InutsukaMiyama92,IwasakiEtAl2011}, of curvature \citep[][]{ChandrasekharFermi53,Ostriker64b,SadhukhanEtAl16}, and, last but not least, of magnetic fields \citep[][]{Strittmatter66,Kellman72,Kellman73,Langer78,NakanoNakamura78,TomisakaIkeuchi83,Nakano88,HosseiniradEtAl17}.
Despite these numerous works, there is still no rigorous and complete study of the magnetized Jeans' instability, i.e. a derivation yielding analytically, without simplifying assumptions, the explicit expressions for eigenvalues and eigenfunctions of the corresponding eigenvalue problem.
The two major difficulties to do so are the following.

The first difficulty is that systems including gravity are necessarily stratified. Considering a homogeneous background violates the static equilibrium Poisson equation, and doing so is known in the literature as Jeans' swindle. The Universe being statistically homogeneous and isotropic at its largest scales and is not static, this simplification yields good results in the cosmological context, aside from studies of the cosmic web, which is obviously a stratified medium.
Now, the study of waves and instabilities in stratified media is a very complex topic, which is still active ongoing research by hydrodynamicists and plasma physicists.
The most celebrated example of an important subtlety arising from inhomogeneity is Landau damping, due to inhomogeneity in velocity space (kinetic description), but a similar damping arises in the fluid description of magnetohydrodynamics (MHD) in spatially inhomogeneous systems.
Mathematically speaking, taking rigorously into account stratification involves considerations on continuous spectra and generalized eigenfunctions (i.e. distributions)
in the framework of spectral theory. The spectral theory of linear operators has a wide field of applications in physics. It is at the foundation of quantum mechanics \citep{VonNeumann55} as well as of MHD \citep{Lifschitz89,GKP}, it is useful in astrophysics \citep{Adam86_PartI,Adam86_PartII,Winfield16} and in Earth seismology \citep{Margerin09_a,Margerin09_b}, to name but a few examples.

The second difficulty is that formally the study of gravitational instability is an eigenvalue problem involving an integro-differential operator (cf. section~\ref{VectorForm} below). Physically, this stems from the fact that gravity is a long range force, without negative masses (mass being the gravitational equivalent of charge in electromagnetism). In Newtonian gravity there is no screening mechanism, no gravitational equivalent of a Debye sphere, which could reduce the interaction to a local one effectively.
Consequently, the system of equations governing this eigenvalue problem is of fourth order.
In some fields, notably in asteroseismology and in laboratory MHD, the perturbation of the gravitational potential induces only small effects, and it is common practice to neglect it. This is called the Cowling approximation \citep{Cox80,UnnoEtAl89}, and doing so reduces this fourth-order problem to a second order one, enabling an approximate analytic treatment. However, this approximation is not relevant for our purpose, since it precisely discards the term responsible for the Jeans instability.

\subsection{Overview of related analytical approaches}

Efforts to understand analytically the evolution of perturbations in self-gravitating structures without the Cowling approximation are ongoing. 
So far, the analytic dispersion relations for Jean's instability were derived in special cases only, notably in the incompressible case
\citep{GoldreichLyndenBell65,Tassoul67}, in the thin sheet limit \citep{TomisakaIkeuchi85,WunschEtAl10}, focusing on marginal stability only \citep{Oganesyan60,GoldreichLyndenBell65}, or working under simplifying assumptions about the scale of perturbations \citep{LubowPringle93,Clarke99}. Variational approaches such as in \cite{Chandrasekhar61,LyndenBellOstriker67,RaoultPellat78} provide general stability criteria but do not give explicit expressions for the eigenvalues. An upper bound on the perturbed self-gravitational energy associated with the Lagrangian displacement was derived by \cite{KeppensDemaerel16_part1,KeppensDemaerel16_part2}, and \cite{DurriveLanger19} decomposed, in the planar hydrostatic case, the fourth-order eigenvalue problem into a sequence of second-order problems that can be solved separately.
In addition, the eigenvalue problem related to Jeans' instability being formally very similar to the one relevant for stellar oscillations, the following analytical studies are also noteworthy.
Since the Cowling approximation is not accurate for long wavelength oscillations \citep{Cox80}, it poorly describes the dipolar f-mode in stellar oscillations, and taking advantage of the fact that dipolar oscillations have the specific property of yielding a first integral from momentum conservation, \cite{Takata05,Takata06} reduced his fourth-order system of equations into a second-order one, and was able to analyze adiabatic dipolar oscillations of stars without making the Cowling approximation.
This analysis is restricted to a specific mode, does not consider magnetic fields, and assumes a hydrostatic equilibrium.
Other mathematically-oriented stellar physics studies include \cite{Beyer95_a,Beyer95_b,BeyerSchmidt95} in the framework of operator theory, and \cite{Takata12} who suggests a way to give a complete mathematical justification to the conventional classification of stellar eigenmodes into p-modes, g-modes, and f-modes, adopting an approach from the field of geoseismology based on wedge products.
Finally, \cite{PoedtsEtAl85} use both the technique from \cite{Goedbloed75} and that from \cite{Pao75} to derive a reduced eigenvalue problem focused on continuum modes. They conclude that the perturbation of the gravitational potential has no effect on the continuous spectrum. However, their study excludes, by construction, discrete modes and in particular Jeans' instability.

\subsection{Aim of this MHD spectral approach}

In the present paper, we aim at addressing the two above difficulties and contributing to the challenge of understanding rigorously the magnetized Jeans' gravitational instability as follows.
Our goal is to reformulate the problem in order to exhibit the fundamental singularities underlying this problem, given that singularities of differential equations are key to understand dynamics \citep[e.g.][]{Adam86_Book}. More precisely, we will extend the approach of \cite{GKP}:
Based on the work of \cite{Goedbloed71}, the modern textbook treatment in \cite{GKP} (section 7.3) exhibits the spectrum of a non-uniformly magnetized plasma slab embedded in a uniform gravitational field, making the Cowling approximation. To do so, they write in Sturm-Liouville form the equation satisfied by the component of the displacement vector in the direction of the stratification. From this equation, the spectrum may be read: the zeros of the numerator of the coefficient of the highest order term correspond to the slow and Alfv\'en genuine singularities (continuous spectra), while the zeros of its denominator correspond to the slow and fast magneto-acoustic apparent singularities. In addition, the theorem derived by \cite{GoedbloedSakanaka74}, which extends the classical Sturm-Liouville oscillation theorem (relevant for a linear eigenvalue problem) to this non-linear eigenvalue problem, indicates the monoticity of the discrete parts of the spectrum lying between these continuous ranges of singularities.
In the present work, we complement this study by deriving the MHD wave equation of a self-gravitating slab, taking into account both the equilibrium and the perturbed Poisson equations.

In the process, we recast the problem into various compact, classical forms, suited to analyze the spectrum and make the solutions explicit. In particular, we manage to factorize this MHD wave equation. Previous authors did not take advantage of the wave equation formulation.
For instance \cite{Ledoux50} and \cite{LedouxWalraven58} say that \cite{Pekeris38} has been the first to carry out completely the necessary eliminations and derived a fourth order equation `which is too complicated to be reproduced [in their paper]'.
Similarly, it is written in \cite{GoldreichLyndenBell65} that
they derived this equation but they `did not find the result very enlightening so [they] shall not repeat it [in their paper]'. 
One example in which such an equation is made fully explicit is \cite{ElmegreenElmegreen78}. However, it is limited to the hydrodynamical and isothermal case, and only the equation on the gravitational potential is derived, while we will formulate the equation on the Lagrangian displacement vector, which is the most fundamental variable since all the other linearized quantities may be directly deduced from it.
In addition, the equation in \cite{ElmegreenElmegreen78} is left unfactorized.
Similarly, often this eigenvalue problem is written as a set of two coupled second order differential equations \citep[e.g. equations (22)-(33) of][]{NagaiEtAl98} or as a $4 \times 4$ matrix differential equation \citep[e.g. equations (15)-(27) of][]{NakamuraEtAl91}, but without any particular form, such that it is impossible to tell from the coefficients which frequencies are the genuine singularities at the heart of the dynamics.

\subsection{Paper organization}

The paper is organized as follows. First, we present the equilibrium state under consideration (section~\ref{section:Equilibrium}), and then the MHD equations linearized about this equilibrium (section~\ref{sec:LinearizedMHD}). In section~\ref{VectorForm} we present the eigenvalue problem, which we then transform into various forms, from which we discuss its spectral properties and its solutions: (i) a coupled Sturm-Liouville form (obtained through sections~\ref{sec:FieldLineProjection}, \ref{sec:4x4MatrixOperator}, \ref{sec:CSL}), (ii) a coupled harmonic oscillator form, including its Hamiltonian form (section~\ref{sec:CHO}), (iii) a first order matrix differential equation (section~\ref{sec:MatrixDifferentialEquation}), and (iv) a scalar wave equation (section~\ref{sec:WaveEquation}).
In section~\ref{sec:FinalStep}, we give the expression of the displacement vector and perturbation of the gravitational potential in terms of the solution of the above matrix differential equation. In that sense, we reduced the challenge of solving the initial problem to solving a much simpler problem (which we do fully solve in a certain limit). In section~\ref{sec:Example} we illustrate through a simple example how we may obtain explicitly the dispersion relation thanks to the above reformulation, and in particular we give the analytic expression for the mode corresponding to Jeans' instability.
Finally, we conclude in section~\ref{sec:Conclusion}, presenting some of the prospects of this work.

\section{Equilibrium relations}
\label{section:Equilibrium}

Let us consider a magnetized, polytropic, self-gravitating, planar medium in static equilibrium, governed by the following relations. Using Cartesian coordinates $x$, $y$ and $z$, we choose $x$ as the direction of stratification. Thus, all equilibrium quantities, denoted with subscripts $0$, are functions of $x$ only. The slab contains a magnetic field $\vec{B}_0$ confined to plane layers perpendicular to the stratified direction $x$, but whose components vary along the stratification, namely
\begin{equation}
\vec{B}_0 = B_y(x) \vec{e}_y + B_z(x) \vec{e}_z,
\end{equation}
where we denote by $\vec{e}_x$, $\vec{e}_y$ and $\vec{e}_z$ the unit vectors in the $x$, $y$ and $z$ directions. As in \cite{GKP}, throughout this paper we make use of units where vacuum permeability $\mu_0$ is unity. To restore mks units one should make the substitutions $\vec{B} \rightarrow \vec{B}/\sqrt{\mu_0}$ and $\vec{j} \rightarrow \sqrt{\mu_0} \vec{j}$ in the formulae. Thus, the equilibrium currents are given by $\vec{j}_0 = \vec{\nabla} \times \vec{B}_0 = - B'_z \vec{e}_y + B'_y \vec{e}_z$. These currents remain along magnetic surfaces (the $(y,z)$ planes at a given height $x$), and since they are in general not aligned with $\vec{B}_0$, there exists a non-vanishing Lorentz force
\begin{equation}
\vec{j}_0 \times \vec{B}_0 = - \vec{\nabla} \left(\tfrac{1}{2} B_0^2\right),
\label{LorentzForce}
\end{equation}
where $B_0^2 = B_y^2+B_z^2$. However, there is no magnetic curvature term in this configuration.
The slab is self-gravitating, meaning that the equilibrium gravitational acceleration $\vec{g}_0$ satisfies the Poisson equation 
\begin{equation}
\vec{\nabla} \cdot \vec{g}_0 = - 4 \pi G \rho_0,
\label{Equilibrium_PoissonVectorial}
\end{equation}
where $\rho_0$ is the equilibrium density field and $G$ is Newton's gravitational constant. This equation introduces the parameter 
\begin{equation}
\omega_0^2 \equiv 4 \pi G \rho_0(x),
\end{equation}
which corresponds physically to the local (due to its $x$-dependence) free-fall timescale. This timescale is the fundamental new ingredient as compared to the textbook treatment in \cite{GKP}. The Lorentz and gravitational forces are in competition with gradients of the pressure $p_0$, such that the equilibrium force balance reads
\begin{equation}
- \vec{\nabla} p_0 + \vec{j}_0 \times \vec{B}_0 + \rho_0 \vec{g}_0 = \vec{0}.
\label{ForceBalance_vectorial}
\end{equation}
Now, given the planar geometry we may write
\begin{equation}
\vec{g}_0 = g_0(x) \vec{e}_x,
\end{equation}
so that Poisson's equation \eqref{Equilibrium_PoissonVectorial} reduces to
\begin{equation}
\displaystyle g_0'=-\omega_0^2,
\label{Equilibrium_Poisson}
\end{equation}
and using \eqref{LorentzForce}, the force balance \eqref{ForceBalance_vectorial} reduces to 
\begin{equation}
\left(p_0 + \tfrac{1}{2} B_0^2 \right)' = \rho_0 g_0.
\label{ForceBalance}
\end{equation}
Throughout this work we consider a polytropic equation of state
\begin{equation}
p_0 = \kappa \rho_0^\gamma,
\label{polytrope}
\end{equation}
where $\gamma$ is called the polytropic exponent and $\kappa$ is a constant that depends on the specific entropy. The isothermal equation of state corresponds to $\gamma =1$, in which case $\kappa$ reduces to the speed of sound squared.
Finally, two local (due to their $x$-dependence) speeds appear in this problem: the speed of sound
\begin{equation}
c(x) \equiv \sqrt{\gamma \frac{p_0}{\rho_0}},
\label{def:c}
\end{equation}
and the Alfv\'en speed
\begin{equation}
b(x) \equiv \frac{B_0}{\sqrt{\rho_0}},
\label{def:b}
\end{equation}
associated with the propagation of purely magnetic waves called Alfv\'en waves, which are in essence vectorial since $\vec{B}_0$ is a vector, but given the planar stratification considered, only the above scalar Alfv\'en speed appears here.

\section{Perturbation equations}
\label{sec:LinearizedMHD}

The ideal MHD equations describing the dynamics of a self-gravitating magnetized fluid are
\begin{equation}
\renewcommand\arraystretch{1.2} 
\begin{array}{l}
\partial_t \rho + \vec{\nabla} \cdot (\rho \vec{v}) = 0,\\
\rho (\partial_t \vec{v} + \vec{v} \cdot \vec{\nabla} \vec{v}) = - \vec{\nabla} p + \vec{j} \times \vec{B} + \rho \vec{g},\\
\vec{j}=\vec{\nabla} \times \vec{B},\\
\partial_t \vec{B} = \vec{\nabla} \times (\vec{v} \times \vec{B}),\\
\vec{\nabla} \cdot \vec{g} = - 4 \pi G \rho,
\end{array}
\label{MHD_eqns}
\end{equation}
corresponding respectively to mass conservation, the momentum equation (with pressure gradients, Lorentz's force and the gravitational force), Amp\`ere's law, the induction equation (Faraday's law with Ohm's law in the infinite electric conductivity limit) and Poisson's equation. To analyze waves and instabilities in self-gravitating magnetized fluids, we linearize these equations around the equilibrium state detailed in section \ref{section:Equilibrium}. Following the usual procedure, for each quantity $Q=(\rho,\vec{v},\vec{B},\vec{j},\vec{g})$ we write $Q = Q_0 + Q_1$ where subscripts $0$ and $1$ indicate equilibrium and perturbed quantities respectively, assuming $|Q_1| \ll |Q_0|$. Doing so, mass conservation and the momentum equation read
\begin{equation}
\partial_t \rho_1 + \vec{\nabla} \cdot \left(\rho_0 \vec{v}_1\right) = 0,
\label{LinearizedMassConservation}
\end{equation}
and
\begin{equation}
\rho_0 \partial_t \vec{v}_1 = - \vec{\nabla} p_1 + \vec{j}_1 \times \vec{B}_0 + \vec{j}_0 \times \vec{B}_1 + \rho_1 \vec{g}_0 + \eta \rho_0 \vec{g}_1,
\label{LinearizedMomentumConservation}
\end{equation}
where $\rho_1, \vec{v}_1,\vec{j}_1,\vec{B}_1$ and $\vec{g}_1$ are respectively the perturbations of mass density, velocity, current density, magnetic field, and gravitational acceleration.
Physically, the terms in the right hand side of \eqref{LinearizedMomentumConservation} correspond to the forces applied on volume elements, and are modeled as follows.
Note that we linearize about a static equilibrium where $\vec{v}_0 = \vec{0}$, such that the linearization of \eqref{MHD_eqns} involves \eqref{LinearizedMassConservation}.

\textit{Gradient of pressure}
The above set of fluid equations requires a closure relation, constraining $p_1$.
Let us consider that the timescales of the perturbations -- the oscillation period if stable and growth timescale if unstable -- are sufficiently short so that no heat is exchanged between neighboring fluid elements. Then the evolution of the perturbations may be considered as adiabatic and, from thermodynamical considerations, it can be shown (cf. \cite{Thompson06} for example) that the equation expressing the absence of heat exchange $\delta Q=0$ becomes the following relation between the Lagrangian variation of pressure $\delta p$ and the Lagrangian variation of density $\delta \rho$:
\begin{equation}
\frac{\delta p}{p_0} = \gamma_\mathrm{ad} \frac{\delta \rho}{\rho_0},
\label{AdiabaticPerturbLagrangian}
\end{equation}
where, in general the constant $\gamma_\mathrm{ad}$ is different from the polytropic exponent $\gamma$ from the polytropic equation of state \eqref{polytrope} of the equilibrium. Let us rewrite equation \eqref{AdiabaticPerturbLagrangian} in the Eulerian variables $\rho_1$ and $p_1$ rather than in Lagrangian variables. The link between the two descriptions is given by
\begin{equation}
\begin{array}{l}
\delta \rho = \rho_1 + \vec{\xi} \cdot \vec{\nabla} \rho_0,\\
\delta p = p_1 + \vec{\xi} \cdot \vec{\nabla} p_0.
\end{array}
\end{equation}
where $\vec{\xi}$ is the Lagrangian displacement vector, which is related to the Eulerian velocity perturbation by $\vec{v}_1=\partial_t \vec{\xi}$, because we start from a static equilibrium ($\vec{v}_0 = \vec{0}$). Defining the adiabatic speed of sound
\begin{equation}
c_\mathrm{ad} \equiv \sqrt{\gamma_\mathrm{ad} \frac{p_0}{\rho_0}},
\end{equation}
which is different from the speed $c$
defined in \eqref{def:c} in the equilibrium state because $\gamma_\mathrm{ad} \neq \gamma$ in general, expression \eqref{AdiabaticPerturbLagrangian} may be rewritten (cf. \cite{Cox80} for example)
\begin{equation}
p_1 = c_\mathrm{ad}^2 \rho_1 + \gamma_\mathrm{ad} \ p_0 \ \vec{\xi} \cdot \vec{A},
\label{ClosureRelationWithBuoyancy}
\end{equation}
where
\begin{equation}
\vec{A} \equiv \frac{\vec{\nabla} \rho_0}{\rho_0} - \frac{\vec{\nabla} p_0}{\gamma_\mathrm{ad} \ p_0} = \left(1 - \frac{\gamma}{\gamma_\mathrm{ad}}\right) \frac{\vec{\nabla} \rho_0}{\rho_0}.
\label{def:A}
\end{equation}
This vector is a well known quantity in stellar physics, and is linked to the Brunt-V\"ais\"al\"a frequency $N$ by the relation $N^2 \equiv -A g_0$. This frequency gives the timescale associated with buoyancy (frequency of oscillations or growth rate of convective instability). The second equality in \eqref{def:A} is valid in the case of a polytrope of exponent $\gamma$, as \eqref{polytrope}, and indicates that stability depends on the ordering between $\gamma$ and $\gamma_\mathrm{ad}$, according to the so-called Schwarzschild criterion: Convective instability occurs when $\gamma>\gamma_\mathrm{ad}$, while the system stably oscillates (g-modes in stars) when $\gamma<\gamma_\mathrm{ad}$.
In the literature, the expression `gravitational instability' is sometimes ambiguous because it may refer to Jeans instability, Rayleigh-Taylor instability or convective instability. Here, our focus is on Jeans' gravitational instability, and therefore, to keep the equations as transparent as possible, we will not take buoyancy into account by considering a convectively neutral medium, i.e. we take $\gamma_\mathrm{ad}=\gamma$. In this case \eqref{ClosureRelationWithBuoyancy} becomes
\begin{equation}
p_1 = c^2 \rho_1,
\label{ClosureRelationNoBuoyancy}
\end{equation}
where $c$ is the equilibrium speed of sound \eqref{def:c}. Relation \eqref{ClosureRelationNoBuoyancy} is our closure relation. Note that, since we are considering a polytrope, this is equivalent to the relation used in \cite{GKP}
\begin{equation}
p_1 = -\vec{\xi} \cdot \vec{\nabla} p_0 - \gamma p_0 \vec{\nabla} \cdot \vec{\xi},
\end{equation}
obtained by linearizing the adiabatic energy equation.

\textit{Lorentz force} The perturbed current density $\vec{j}_1$ is given by the linearized Amp\`ere law
\begin{equation}
\vec{j}_1 = \vec{\nabla} \times \vec{B}_1,
\end{equation}
where the magnetic field perturbation $\vec{B}_1$ satisfies the linearized induction equation
\begin{equation}
\partial_t \vec{B}_1 = \vec{\nabla} \times \left(\vec{v}_1 \times \vec{B}_0\right).
\end{equation}

\textit{Gravity} The perturbation of the gravitational acceleration $\vec{g}_1$ satisfies the linearized Poisson equation
\begin{equation}
\vec{\nabla} \cdot \vec{g}_1 = - 4 \pi G \rho_1.
\label{LinearizedPoissonEquation}
\end{equation}
Note that the gravitational acceleration $\vec{g}_1$ is a vectorial quantity while the above Poisson equation is only a scalar relation, so that it is not constraining enough to define $\vec{g}_1$ fully. To keep the same amount of information as in Poisson's equation $\Delta \phi_1 = 4 \pi G \rho_1$ for the gravitational potential $\phi_1$, we must add the constraint
\begin{equation}
\vec{\nabla} \times \vec{g}_1 = \vec{0},
\label{constraint_g1}
\end{equation}
which stems from the fact that the gravitational acceleration is a gradient ($\vec{g}_1 = - \vec{\nabla} \phi_1$). The vector relation \eqref{constraint_g1} seems to introduce three constraints, i.e. one more than needed, but in fact one of them is redundant with the others, so that \eqref{constraint_g1} does fix coherently the two degrees of freedom left in \eqref{LinearizedPoissonEquation} to define $\vec{g}_1$ fully.
A crucial feature of the linearized Poisson equation \eqref{LinearizedPoissonEquation} is that it does not give $\vec{g}_1$ explicitly, but it only fixes its divergence. This information may be recast in integral form, omitting surface terms, as
\begin{equation}
\vec{g}_1 = - G \int \rho_1(\vec{r'}) \frac{\vec{r} - \vec{r}'}{|\vec{r} - \vec{r}'|^3} d^3r',
\label{IntegralForm_g1}
\end{equation}
which exhibits the non-local nature of gravity, with important consequences detailed in the next section.
In the momentum equation \eqref{LinearizedMomentumConservation}, the two last terms are due to gravity, but they act in two different ways. Hereafter we will call the first gravity term, $\rho_1 \vec{g}_0$, the `Cowling term' and the other one the `Jeans term'. The Jeans term is the additional term that we consider compared to \cite{GKP}. The customary way to write this term is $\rho_0 \vec{g}_1$ but we introduce in front of it a parameter $\eta$, that we call the `gravitational dilution factor' following \cite{ChristensenDalsgaardGough01} who also introduced it. This parameter will help us keep track of the impact of this additional term throughout our calculations, and will facilitate comparisons between our results and the literature. Indeed, setting
\begin{equation}
\eta = 0,
\label{CowlingApprox}
\end{equation}
is known in the literature as the `Cowling approximation'. It is always relevant in laboratory MHD and is often appropriate in asteroseismology (\cite{Cox80,SmeyersVanHoolst10} for example). The Cowling term encodes the effect of the unperturbed gravitational field on the density perturbations, which may give rise to the Rayleigh-Taylor instability and to convection when buoyancy is included. On the contrary, setting
\begin{equation}
\eta = 1,
\end{equation}
corresponds to the full description of gravity, in which case the Jeans term reads $\rho_0 \vec{g}_1$ as it should. The Jeans term is the term which may give rise to the Jeans gravitational instability, and gravitational fragmentation. It is therefore essential in the context of star formation and galaxy formation. The counterpart of the Cowling approximation is called the `Jeans swindle'. It consists in considering a homogeneous equilibrium density $\rho_0$, so that the Cowling term is absent.
Naturally, this assumption has the advantage of immensely simplifying the problem since one may then simply consider plane waves in all directions. Now, although drastic, this simplification provides good predictions in some cases, notably at the largest cosmological scales at which the Universe is indeed homogeneous and isotropic, statistically speaking, and the background is not static. However, taking a uniform $\rho_0$ violates the static equilibrium Poisson equation, and this `swindle' is not enough to make correct predictions in many contexts, since in general the density stratification does play an important dynamical role. Therefore, the purpose of our work is to analyze the above linearized MHD equations taking gravity fully into account, i.e. without making neither the Cowling approximation nor the Jeans swindle, and track all waves and instabilities, particularly the Jeans instability, in self-gravitating stratified magnetized media.

\section{Eigenvalue problem in vector form}
\label{VectorForm}

Let us now formulate the above linearized MHD equations as an eigenvalue problem.
The equilibrium being static, the Eulerian velocity perturbation $\vec{v}_1$ and the Lagrangian displacement vector $\vec{\xi}$ are simply related by $\vec{v}_1=\partial_t \vec{\xi}$.
Also, since the equilibrium quantities do not depend on time (the case of a time-dependent background with self-gravity was analysed in \cite{KeppensDemaerel16_part1}) we may consider solutions in the form of normal modes
\begin{equation}
\vec{\xi}(t,\vec{x}) = \hat{\vec{\xi}}(\vec{x}) e^{-i \omega t}.
\label{NormalModes}
\end{equation}
For convenience, we will drop the hat in the $\hat{\vec{\xi}}$ notation from now on. Then the linearized momentum equation \eqref{LinearizedMomentumConservation} becomes the vector eigenvalue problem (of eigenparameter $\omega^2$ and eigenfunction $\vec{\xi}$)
\begin{equation}
- \omega^2 \rho_0 \vec{\xi} = \vec{F}\left(\vec{\xi}\right),
\label{VectorEigenvalueProblem}
\end{equation}
where $\vec{F}$ is an operator acting on $\vec{\xi}$, called the `force operator', with
\begin{equation}
\vec{F}(\vec{\xi}) = - \vec{\nabla} p_1 + \vec{j}_1 \times \vec{B}_0 + \vec{j}_0 \times \vec{B}_1 + \rho_1 \vec{g}_0 + \eta \rho_0 \vec{g}_1.
\label{ForceOperator}
\end{equation}
For a thorough presentation of this operator, in full generality, see \cite{GKP,KeppensDemaerel16_part1}. A key feature is that the operator $\rho^{-1} \vec{F}$ is self-adjoint\footnote{At least under appropriate boundary conditions. Indeed, as shown for instance in \cite{GKP,KeppensDemaerel16_part1}, demonstrating the self-adjointness of the force operator (and not only its symmetry) involves integrations by parts (the integrals stemming from the definition of the inner product) which introduce surface integrals. For self-adjointness to hold, the latter are required to vanish, which is possible under appropriate boundary or symmetry conditions, assumed to be adopted here. However, the derivation presented in our paper is independent of this fact, since boundary conditions come into play only at the very end, as in the example of section \ref{sec:Example}.}, which guarantees that we only have waves ($\omega^2>0$) or instabilities ($\omega^2<0$) and all instabilities must go through the marginal frequency ($\omega^2=0$).

The variables $\rho_1, p_1, \vec{B}_1, \vec{j}_1$ and $\vec{g}_1$ in \eqref{ForceOperator} are seen as operators acting on $\vec{\xi}$, with the following expressions. The linearized mass conservation gives the density perturbation
\begin{equation}
\rho_1 = - \vec{\nabla} \cdot \left(\rho_0 \vec{\xi}\right),
\label{AsAFunctionOfXi_rho1}
\end{equation}
our choice of closure relation \eqref{ClosureRelationNoBuoyancy} gives the pressure perturbation
\begin{equation}
p_1 = - c^2 \vec{\nabla} \cdot \left(\rho_0 \vec{\xi}\right),
\label{AsAFunctionOfXi_p1}
\end{equation}
the induction equation gives the magnetic field perturbation
\begin{equation}
\vec{B}_1 = \vec{\nabla} \times \left(\vec{\xi} \times \vec{B}_0\right),
\label{AsAFunctionOfXi_B1}
\end{equation}
Amp\`ere's law gives the current density perturbation
\begin{equation}
\vec{j}_1 = \vec{\nabla} \times \left(\vec{\nabla} \times \left(\vec{\xi} \times \vec{B}_0\right)\right),
\label{AsAFunctionOfXi_j1}
\end{equation}
and finally, the integral form of Poisson's equation \eqref{IntegralForm_g1} gives the gravitational acceleration perturbation
\begin{equation}
\vec{g}_1 = G \int \dv \left(\rho_0 \vec{\xi}\right) \frac{\vec{r} - \vec{r}'}{|\vec{r} - \vec{r}'|^3} d^3r' \ .
\label{AsAFunctionOfXi_g1}
\end{equation}
Expression \eqref{AsAFunctionOfXi_g1} reveals the fundamental feature introduced by the Jeans term: in the Cowling approximation the force operator is differential, while here we have to deal with an integro-differential operator.
Now, following \cite{GKP}, our aim is to derive the scalar wave equation, i.e. the scalar equation satisfied by the component of $\vec{\xi}$ in the direction of the stratification. As we will see, because of this integral part in the vector eigenvalue equation \eqref{VectorEigenvalueProblem}, we will end up with a differential equation of the fourth order, while it is of the second order in \cite{GKP}. Therefore, the Cowling approximation reduces the order of the wave equation, and this eliminates potentially important information: (i) two linearly independent fundamental solutions out of four are discarded, (ii) two boundary conditions out of four are neglected, and (iii) the coefficient in front of the highest order term of the wave equation, which is the key feature to determine the spectrum as demonstrated by \cite{GKP}, is to be re-examined, since already the order itself of this highest order term is modified. For example a singularity in the Cowling approximation may not be a singularity in the full wave equation. The purpose of the gravitational dilution factor $\eta$ that we introduced in the force operator is precisely to assess the impact of the Cowling approximation on the wave equation.
Finally, one may argue that we lost some generality because we neglected surface terms in expression \eqref{AsAFunctionOfXi_g1} of $\vec{g}_1$. However, in the following we do not use \eqref{AsAFunctionOfXi_g1}: we listed this expression to highlight how different the Jeans term is compared to the other terms in the force operator. Instead of using \eqref{AsAFunctionOfXi_g1}, in the following derivation we incorporate $\vec{g}_1$ exclusively through the differential expressions \eqref{LinearizedPoissonEquation} and \eqref{constraint_g1}, so that our results do not depend on a particular choice of boundary conditions (except at the very end, in section~\ref{sec:Example}, where we add boundary conditions explicitly).

\section{Field line projection}
\label{sec:FieldLineProjection}

In order to facilitate the derivation of the scalar wave equation, let us introduce a couple of definitions. Thanks to the translation invariance in the $(y,z)$ plane of the equilibrium state, we Fourier transform in the $y$ and $z$ directions, and can consider, without loss of generality,
\begin{equation}
\vec{\xi} = \left[\xi_x(x) \ \vec{e}_x + \xi_y(x) \ \vec{e}_y + \xi_z(x) \ \vec{e}_z\right] e^{i (k_y y + k_z z)}.
\label{xi_MHD}
\end{equation}
Similarly we put\footnote{Given \eqref{AsAFunctionOfXi_g1} we take the same time dependence as $\vec{\xi}$ in~\eqref{NormalModes}. Also, to be precise, we follow the same steps as with $\vec{\xi}$, namely we set $\vec{g}_1(t,\vec{x}) = \hat{\vec{g}}_1(\vec{x}) e^{-i \omega t}$, with $\hat{\vec{g}}_1(\vec{x})$ given by \eqref{def_g1}, but for convenience we get rid of the hat on $\hat{\vec{g}}_1$.}
\begin{equation}
\vec{g}_1 = \left[g_{1x} (x) \vec{e}_x + g_{1y} (x) \vec{e}_y + g_{1z} (x) \vec{e}_z\right] e^{i (k_y y + k_z z)}.
\label{def_g1}
\end{equation}
In addition, we work in the field line projection, i.e. we define the field line triad
\begin{equation}
\begin{array}{lclcl}
\vec{e}_x & \equiv & \vec{\nabla} x, & & \\
\vec{e}_\perp & \equiv & \vec{B}_0/B_0 \times \vec{e}_x & = & (B_z \vec{e}_y - B_y \vec{e}_z)/B_0,\\
\vec{e}_\parallel & \equiv & \vec{B}_0/B_0 & = & (B_y \vec{e}_y + B_z \vec{e}_z)/B_0,\\
\end{array}
\end{equation}
such that the gradient operator reads
\begin{equation}
\vec{\nabla} = \vec{e}_x \partial_x + i \vec{e}_\perp(x) k_\perp + i \vec{e}_\parallel(x) k_\parallel,
\label{gradientOperator}
\end{equation}
where $k_\perp$ and $k_\parallel$ represent the perpendicular and parallel derivatives
\begin{equation}
\begin{array}{lclcl}
k_\perp(x) & \equiv & - i \vec{e}_\perp \cdot \vec{\nabla} & = & (k_y B_z - k_z B_y)/B_0,\\
k_\parallel(x) & \equiv & - i \vec{e}_\parallel \cdot \vec{\nabla} & = & (k_y B_y + k_z B_z)/B_0,
\end{array}
\end{equation}
where we used the substitutions $\partial_y \rightarrow i k_y$ and $\partial_z \rightarrow i k_z$ for the invariant directions. The functions $k_\perp$ and $k_\parallel$ may be considered as the wave vectors in the perpendicular and parallel directions, respectively. They are $x$-dependent but the resulting horizontal wave vector $k_0$ is not:
\begin{equation}
k_0 \equiv \sqrt{k_\perp^2+k_\parallel^2} = \sqrt{k_y^2+k_z^2} = \text{constant}.
\end{equation}
In this projection we have
\begin{equation}
\renewcommand\arraystretch{1.2} 
\begin{array}{rcl}
\vec{\xi} \! \! & = & \hspace{-0.22cm} \left( \ \xi \ \vec{e}_x \! \! \ - i \xi_\perp \ \vec{e}_\perp - i \xi_\parallel \ \vec{e}_\parallel\right) e^{i (k_y y + k_z z)},\\
\vec{g}_1 \hspace{-0.25cm} & = & \hspace{-0.25cm} \left(g_x \ \vec{e}_x - i g_\perp \ \vec{e}_\perp - i g_\parallel \ \vec{e}_\parallel\right) e^{i (k_y y + k_z z)},
\end{array}
\label{FieldLineProjection_of_xi_and_g1}
\end{equation}
where the components are defined as
\begin{equation}
\begin{array}{l}
\renewcommand\arraystretch{1.2} 
\left( \! \!
\begin{array}{c}
\xi\\
\xi_\perp \\
\xi_\parallel
\end{array}
 \! \! \right)
\equiv
\left( \! \!
\begin{array}{c}
\xi_x\\
i \vec{e}_\perp \cdot \vec{\xi}\\
i \vec{e}_\parallel \cdot \vec{\xi}
\end{array}
 \! \! \right)
=
\left( \! \!
\begin{array}{c}
\xi_x\\
i (B_z \xi_y - B_y \xi_z)/B_0\\
i (B_y \xi_y + B_z \xi_z)/B_0
\end{array}
 \! \! \right),
\end{array}
\label{def:xi_components}
\end{equation}

\vspace{2cm}

\noindent
and
\begin{equation}
\hspace{-0.4cm}
\begin{array}{l}
\renewcommand\arraystretch{1.2} 
\left( \! \!
\begin{array}{c}
g_x\\
g_\perp \\
g_\parallel
\end{array}
 \! \! \right)
\equiv
\left( \! \!
\begin{array}{c}
g_{1x}\\
i \vec{e}_\perp \cdot \vec{g}_1\\
i \vec{e}_\parallel \cdot \vec{g}_1
\end{array}
 \! \! \right)
=
\left( \! \!
\begin{array}{c}
g_{1x}\\
i (B_z g_{1y} - B_y g_{1z})/B_0\\
i (B_y g_{1y} + B_z g_{1z})/B_0
\end{array}
 \! \! \right).
\end{array}
\label{def:g_components}
\end{equation}
As in \cite{GKP} we have inserted factors $i$ because this turns out to lead to a representation where $\xi, \xi_\perp$ 	and $ \xi_\parallel$ may be assumed to be real.

\section{Matrix Operator form}
\label{sec:4x4MatrixOperator}

Let us write the spectral equation \eqref{VectorEigenvalueProblem} in the field line projection \eqref{FieldLineProjection_of_xi_and_g1}. This step only consists in straightforward but rather tedious calculations. Ultimately, because of our additional term $\vec{g}_1$ in the force operator, we obtain a generalization of the matrix representation of the spectral equation (7.78) of \cite{GKP}, namely
\begin{equation}
\mathsf{F} \cdot \textbf{X} + \eta \rho_0 \textbf{Y} = - \rho_0 \omega^2 \textbf{X},
\label{matrixRepresentation_Intermediate}
\end{equation}
where
\begin{equation}
\textbf{X} \equiv (\xi,\xi_\perp,\xi_\parallel)^\textsc{T}, \hspace{0.3cm} \textbf{Y} \equiv (g_x,g_\perp,g_\parallel)^\textsc{T},
\end{equation}
and
\begin{widetext}
\begin{equation}
\mathsf{F} \equiv
\left(
\renewcommand\arraystretch{1.4} 
\begin{array}{ccc}
\frac{\dd}{\dd x} \rho_0 (c^2 \! + \! b^2) \frac{\dd}{\dd x} -k_\parallel^2 \rho_0 b^2 + \rho_0 g_0' & \frac{\dd}{\dd x} k_\perp \rho_0 (c^2 \! + \! b^2) - k_\perp \rho_0 g_0 & \frac{\dd}{\dd x} \rho_0 c^2 k_\parallel - k_\parallel \rho_0 g_0 \\
- k_\perp \rho_0 (c^2 \! + \! b^2) \frac{\dd}{\dd x} - k_\perp \rho_0 g_0 & - k_\perp^2 \rho_0 (c^2 \! + \! b^2) - k_\parallel^2 \rho_0 b^2 & - k_\parallel k_\perp \rho_0 c^2 \\
- k_\parallel \rho_0 c^2 \frac{\dd}{\dd x} - k_\parallel \rho_0 g_0 & - k_\parallel k_\perp \rho_0 c^2 & - k_\parallel^2 \rho_0 c^2
\end{array}
\right),
\label{MatrixF}
\end{equation}
\end{widetext}
where we have used the equilibrium force balance \eqref{ForceBalance} to simplify the expression.
Our expression of $\mathsf{F}$ differs from (7.78) of \cite{GKP} in two ways: we have a sign difference in front of $g_0$ and we have an additional term $\rho_0 g_0'$ in the first row. The first difference is simply a matter of convention, as we defined $\vec{g}_0 \equiv g_0 \vec{e}_x$ while they introduce a negative sign in this definition, and the second difference comes from the fact that in (7.78) of \cite{GKP} the gravitational acceleration is assumed to be a constant while here, since we consider a self-gravitating slab, it is a function of $x$. Besides these differences, when setting $\eta=0$ in \eqref{matrixRepresentation_Intermediate} we recover (7.78) of \cite{GKP}, as we should.
Equation \eqref{matrixRepresentation_Intermediate} can also be viewed as the planar version of (13.80) of \cite{GKP} \citep[in which $g_0$ depends on the radial coordinate, see also][]{KeppensEtAl02} where we added self-gravity.

The representation \eqref{matrixRepresentation_Intermediate} involves the six variables $(\xi,\xi_\perp,\xi_\parallel,g_x,g_\perp,g_\parallel)$. Let us now reduce this description to four variables, using the constraints satisfied by $\vec{g}_1$. Firstly, in the field line projection the constraint \eqref{constraint_g1}, that states that $\vec{g}_1$ derives from a potential, reads
\begin{equation}
\renewcommand\arraystretch{1.2} 
\begin{array}{l}
\displaystyle g_\parallel = \frac{k_\parallel}{k_\perp} g_\perp,\\
\displaystyle g_{x} = - \left(\frac{k_\perp g_\perp + k_\parallel g_\parallel}{k_0^2}\right)'.
\end{array}
\label{FieldLineProjection_of_irrotationalConstraint}
\end{equation}
This second relation suggests to define the variable
\begin{equation}
\mathcal{G} \equiv \frac{k_\perp g_\perp + k_\parallel g_\parallel}{k_0^2}.
\label{def:variableG}
\end{equation}
It is not surprising that a variable of this form turns out to be convenient for the present problem, because the equilibrium state is invariant in the plane perpendicular to the stratification direction~$x$ (cf. the form \eqref{gradientOperator} of the gradient operator). Secondly, in the above variables and projection, the Poisson equation \eqref{LinearizedPoissonEquation} for the perturbed quantities becomes
\begin{equation}
\mathcal{G}'' - k_0^2 \mathcal{G} = - (\omega_0^2 \xi)' - \omega_0^2 (k_\perp \xi_\perp + k_\parallel \xi_\parallel).
\label{FieldLineProjection_of_PoissonEquation}
\end{equation}
Therefore, our eigenvalue problem \eqref{VectorEigenvalueProblem} can be put into the second-order $4\times4$ matrix operator form
\begin{equation}
\arraycolsep=1.4pt\def\arraystretch{2.2} 
\left(
\renewcommand\arraystretch{1.4} 
\begin{array}{cccc}
\frac{\dd^2}{\dd x^2} - k_0^2 & \frac{\dd}{\dd x} \omega_0^2 & \omega_0^2 k_\perp & \omega_0^2 k_\parallel \\
- \eta \rho_0 \frac{\dd}{\dd x} & \rho_0 \omega^2 + \mathsf{F}_{11} & \mathsf{F}_{12} & \mathsf{F}_{13} \\
\eta \rho_0 k_\perp & \mathsf{F}_{21} & \rho_0 \omega^2 + \mathsf{F}_{22} & \mathsf{F}_{23} \\
\eta \rho_0 k_\parallel & \mathsf{F}_{31} & \mathsf{F}_{32} & \rho_0 \omega^2 + \mathsf{F}_{33}
\end{array}
\right)
\renewcommand\arraystretch{1.3} 
\left(
\begin{array}{c}
\mathcal{G} \\
\xi \\
\xi_\perp \\
\xi_\parallel
\end{array}
\right)
=
\vec{0},
\label{MatrixOperatorForm}
\end{equation}
where we denote by $\mathsf{F}_{ij}$ the coefficients of matrix $\mathsf{F}$ given by \eqref{MatrixF}. To derive this, we have also used the first relation of \eqref{FieldLineProjection_of_irrotationalConstraint} which, with \eqref{def:variableG}, gives
\begin{equation}
g_\perp = k_\perp \mathcal{G}, \hspace{0.3cm} \text{and} \hspace{0.3cm} g_\parallel = k_\parallel \mathcal{G}.
\label{gperp_gpara}
\end{equation}
The first line of \eqref{MatrixOperatorForm} corresponds to the linearized Poisson equation and the three others to the linearized momentum equation \eqref{VectorEigenvalueProblem}.
In \eqref{MatrixOperatorForm} the eigenparameter $\omega$, which is to be determined, appears only in the bottom $3\times3$ submatrix. Moreover, since the $\mathsf{F}_{ij}$'s do not contain any $\omega$, we can see that the eigenparameter appears exclusively squared. This means that there are six waves to be found in this description, which appear as three pairs of forward-backward modes (this forward-backward symmetry comes from the fact that the equilibrium is static, as it is broken by equilibrium flows). They relate to the slow, Alfv\'en and fast magneto-acoustic mode pairs. 
Finally, we can see that setting $\eta = 0$ decouples this $4\times4$ system into the $3 \times 3$ system containing matrix $\mathsf{F}$, plus an additional equation which then looses its physical meaning of Poisson's equation since the Cowling approximation ($\eta=0$) is not a rigorous, physically based, approximation.

\section{Coupled Sturm-Liouville form}
\label{sec:CSL}

To further explore the dynamics, let us rewrite the two first lines of \eqref{MatrixOperatorForm} as
\begin{equation}
\begin{array}{l}
\hspace{-1cm}
\left(
\begin{array}{cc}
\frac{\dd^2}{\dd x^2} - k_0^2 & \frac{\dd}{\dd x} \omega_0^2\\
- \eta \rho_0 \frac{\dd}{\dd x} & \rho_0 \omega^2 + \mathsf{F}_{11}\\
\end{array}
\right)
\left(
\begin{array}{c}
\mathcal{G} \\
\xi
\end{array}
\right)
\\
\hspace{-1cm}
+
\left(
\begin{array}{cc}
\omega_0^2 k_\perp & \omega_0^2 k_\parallel\\
\mathsf{F}_{12} & \mathsf{F}_{13}\\
\end{array}
\right)
\left(
\begin{array}{c}
\xi_\perp \\
\xi_\parallel
\end{array}
\right)
=
\vec{0},
\end{array}
\label{TwoFirstLines}
\end{equation}
and its two last lines as
\begin{equation}
\begin{array}{l}
\left(
\begin{array}{cc}
\eta \rho_0 k_\perp & \mathsf{F}_{21}\\
\eta \rho_0 k_\parallel & \mathsf{F}_{31}\\
\end{array}
\right)
\left(
\begin{array}{c}
\mathcal{G} \\
\xi
\end{array}
\right)
\\
+
\left(
\begin{array}{cc}
\rho_0 \omega^2 + \mathsf{F}_{22} & \mathsf{F}_{23}\\
\mathsf{F}_{32} & \rho_0 \omega^2 + \mathsf{F}_{33}\\
\end{array}
\right)
\left(
\begin{array}{c}
\xi_\perp \\
\xi_\parallel
\end{array}
\right)
=
\vec{0}.
\end{array}
\label{TwoLastLines}
\end{equation}
In the following calculations, to easily keep track of the order of the derivatives, it is convenient to notice that each $\mathsf{F}_{ij}$ is an operator of order equal to the number of its indices equal to one: for example $\mathsf{F}_{21}$ is a first order operator (the notation $\mathsf{F}_{21}$ contains a single index $1$), while $\mathsf{F}_{11}$ is of second order. This simple rule highlights the important fact that the coefficients in the second matrix in \eqref{TwoLastLines} are algebraic. As a result, we can easily express $\xi_\perp$ and $\xi_\parallel$ in terms of $\xi$ and $\mathcal{G}$ by simply inverting this $2\times2$ matrix. Hence
\begin{equation}
\begin{array}{l}
\left(
\begin{array}{c}
\xi_\perp \\
\xi_\parallel
\end{array}
\right)
=
-\frac{1}{\rho_0^2 D}
\left(
\begin{array}{cc}
\rho_0 \omega^2 + \mathsf{F}_{33} & -\mathsf{F}_{23}\\
-\mathsf{F}_{32} & \rho_0 \omega^2 + \mathsf{F}_{22}\\
\end{array}
\right)
\\
\hspace{1.5cm}
\times \left(
\begin{array}{cc}
\eta \rho_0 k_\perp & \mathsf{F}_{21}\\
\eta \rho_0 k_\parallel & \mathsf{F}_{31}\\
\end{array}
\right)
\left(
\begin{array}{c}
\mathcal{G} \\
\xi
\end{array}
\right),
\end{array}
\label{XiPerp_XiParallel}
\end{equation}
where inverting the matrix introduces the determinant $D \equiv (\omega^2 + \mathsf{F}_{22}/\rho_0) (\omega^2 + \mathsf{F}_{33}/\rho_0) - \mathsf{F}_{23} \mathsf{F}_{32}/\rho_0^2$, which explicitly reads
\begin{equation}
D = \omega^4 - k_0^2 (b^2+c^2) \omega^2 + k_0^2 k_\parallel^2 b^2 c^2.
\label{def:D}
\end{equation}
Defining the slow and fast turning point frequencies
\begin{equation}
\renewcommand\arraystretch{2.7} 
\begin{array}{c}
\displaystyle
\omega_{s0}^2 \equiv \frac{1}{2} k_0^2 (b^2+c^2) \left[1 - \sqrt{1- \frac{4 k_\parallel^2 b^2 c^2}{k_0^2 (b^2+c^2)^2} }\right],\\
\displaystyle
\omega_{f0}^2 \equiv \frac{1}{2} k_0^2 (b^2+c^2) \left[1 + \sqrt{1- \frac{4 k_\parallel^2 b^2 c^2}{k_0^2 (b^2+c^2)^2} }\right],
\end{array}
\end{equation}
this is more conveniently written
\begin{equation}
D=(\omega^2-\omega_{s0}^2)(\omega^2-\omega_{f0}^2).
\label{Def:D}
\end{equation}
Then injecting \eqref{XiPerp_XiParallel} into \eqref{TwoFirstLines} we get a second order system coupling $\xi$ and $\mathcal{G}$.
Making explicit the coefficients, with the definition \eqref{MatrixF} of $\mathsf{F}$ and \eqref{def:D} of $D$, gives
\begin{subequations}
	\begin{align}
		\begin{split}
			\mathcal{G}'' -k_0^2 \left(1 + \eta \tfrac{\omega_\star^2}{\omega^2}\right) \mathcal{G} + \omega_\star^2 \left(\xi' + k_\star \xi\right) = 0,
		\label{2ndOrderSystem_Poisson}
		\end{split}\\
		\begin{split}
			\frac{\eta}{4 \pi G} \left[\left(\omega_\star^2 \mathcal{G}\right)' - \omega_\star^2 k_\star \ \! \mathcal{G}\right] - (P \xi')'-Q \xi= 0,
		\label{2ndOrderSystem_MomentumConservation}
		\end{split}
	\end{align}
\label{2ndOrderSystem}
\end{subequations}
where
\begin{equation}
\renewcommand\arraystretch{1.2} 
\begin{array}{l}
P \equiv N/D,\\
N \equiv \rho_0 (b^2+c^2) (\omega^2-\omega_A^2)(\omega^2-\omega_S^2),\\
Q \equiv \rho_0 (\omega^2-k_\parallel^2 b^2)+\rho_0' g_0-k_0^2 \rho_0 g_0^2 (\omega^2-k_\parallel^2 b^2)/D\\
\hspace{0.7cm} -\left[\rho_0 g_0 \omega^2 (\omega^2-k_\parallel^2 b^2)/D\right]',
\end{array}
\label{CoeffsMHD_Cowling}
\end{equation}
with the Alfv\'en and slow continuum frequencies
\begin{equation}
\renewcommand\arraystretch{1.2} 
\begin{array}{l}
\omega_A^2 \equiv k_\parallel^2 b^2,\\
\omega_S^2 \equiv k_\parallel^2 \frac{b^2 c^2}{b^2 + c^2}.
\end{array}
\label{def:AlfvenSlowContinua}
\end{equation}
The above notations correspond\footnote{Except that we use opposite sign conventions for $Q$ and $g_0$, and our $g_0$ is not assumed to be constant.} to that of \cite{GKP}, but here we in addition introduce the squared pulsation and the wavenumber
\begin{subequations}
	\begin{empheq}{align}
     & \omega_\star^2 \equiv \frac{\omega^2 \omega_0^2 (\omega^2-\omega_A^2)}{D},
        \label{def:omegastar}\\
     & k_\star \equiv \frac{(\omega_0^2)'}{\omega_\star^2} + \frac{k_0^2 g_0}{\omega^2},
        \label{def:kstar}
    \end{empheq}
\label{def:omegastar_and_kstar}
\end{subequations}
which explicitly read
\begin{subequations}
	\begin{empheq}{align}
     & \omega_\star^2 = \frac{\omega^2 \omega_0^2 (\omega^2-\omega_A^2)}{(\omega^2 - \omega_{s0}^2)(\omega^2 - \omega_{f0}^2)},
        \label{def:omegastar_Explicit}\\
     & k_\star = \frac{1}{\omega^2} \left[ \frac{(\omega_0^2)'}{\omega_0^2} \frac{(\omega^2 - \omega_{s0}^2)(\omega^2 - \omega_{f0}^2)}{(\omega^2-\omega_A^2)} + k_0^2 g_0\right].
        \label{def:kstar_Explicit}
    \end{empheq}
\label{def:omegastar_and_kstar_Explicit}
\end{subequations}
In essence, \eqref{2ndOrderSystem_Poisson} corresponds to the linearized Poisson equation, and \eqref{2ndOrderSystem_MomentumConservation} comes from the momentum equation.
It is therefore natural that taking $\eta = 0$ in \eqref{2ndOrderSystem} the two equations decouple, since in the Cowling approximation the perturbed Poisson equation becomes irrelevant.

Let us comment further on this system in the Cowling approximation. When $\eta = 0$ relation \eqref{2ndOrderSystem_MomentumConservation} becomes
\begin{equation}
(P \xi')'+Q \xi= 0,
\label{WaveEquation_CowlingApprox}
\end{equation}
i.e. we recover, as we should, the wave equation derived by \cite{GKP}.
Now, \cite{GKP} analyze \eqref{WaveEquation_CowlingApprox} as follows. Firstly, the force operator \eqref{ForceOperator} being self-adjoint, the eigenvalue $\omega^2$ must be a real number.
In terms of $\omega$, the spectrum thus lies on either the real frequency axis, or the imaginary frequency axis, rather than spanning the full complex plane.
Secondly, \eqref{WaveEquation_CowlingApprox} is a singular differential equation in two different ways: for a given frequency $\omega^2$, there may be positions $x_0$ at which (i) $P$ vanishes, i.e. such that $N(x_0;\omega^2)=0$, and (ii) $P$ diverges, i.e. such that $D(x_0;\omega^2)=0$. As detailed in \cite{GKP}, the $N=0$ singularities give rise to non-square integrable solutions associated with continuous spectra, while the $D=0$ singularities, which seem genuine, are in fact only apparent. Indeed, it turns out that the coefficients \eqref{CoeffsMHD_Cowling} of the equation \eqref{WaveEquation_CowlingApprox} satisfy a special relation which makes the solutions remain finite when $D \rightarrow 0$, i.e. as $\omega^2$ approaches $\omega_{f0}^2$ or $\omega_{s0}^2$.
Hence, the ranges of frequencies corresponding to $N=0$, namely $\{\omega_{A}^2\}$ and $\{\omega_{S}^2\}$, are called continua, while the ranges corresponding to $D=0$, namely $\{\omega_{f0}^2\}$ and $\{\omega_{s0}^2\}$, are not part of the spectrum, and are called ranges of cutoff or turning point frequencies.
Thirdly, a theorem, demonstrated by \cite{GoedbloedSakanaka74}, completes this picture of the structure of the spectrum. To understand its result, let us recall the classical Sturm-Liouville oscillation theorem: in an eigenvalue problem governed by an equation of the form \eqref{WaveEquation_CowlingApprox} but where $P$ is independent of the eigenvalue and $Q$ depends only linearly on the eigenvalue (and under specific boundary conditions), this theorem states that the larger the eigenvalue, the faster the eigenfunction oscillates, i.e. the more nodes it possesses. Such a behavior is called Sturmian. Conversely, in problems where the eigenfunction oscillates more slowly for larger eigenvalues, the behavior is called anti-Sturmian. Now in \eqref{WaveEquation_CowlingApprox}, the eigenparameter $\omega^2$ appears non-linearly in $P$ and in $Q$, so that the ideal MHD eigenvalue problem is often referred to as a non-linear Sturm-Liouville problem. \cite{GoedbloedSakanaka74} have generalized Sturm's oscillation theorem for this MHD wave equation \eqref{WaveEquation_CowlingApprox}. The result is that outside the ranges $\{\omega_{A}^2\}$, $\{\omega_{S}^2\}$, $\{\omega_{f0}^2\}$ and $\{\omega_{s0}^2\}$, the spectrum is discrete with Sturmian behavior for $P>0$ and anti-Sturmian behavior for $P<0$. Therefore, the monoticity of the discrete spectrum changes every time $\omega^2$ crosses one of the four aforementioned ranges. For completeness, let us mention that a fast magneto-sonic point spectrum accumulates at infinity. Another important property, which is demonstrated under specific boundary conditions in \cite{GKP}, is that the eigenfunctions of the discrete spectrum form an orthogonal set.

All these results belong to the wide field of spectral theory in mathematics, and in the context of laboratory plasma experiments like tokamaks, this approach gave rise to MHD spectroscopy. Our aim is to find out how these results generalize once self-gravity is fully taken into account.
To this end, the first step we suggest is to rewrite the system \eqref{2ndOrderSystem} into a classical form, that we link to the mathematical literature on fourth order differential systems. Indeed, while \eqref{WaveEquation_CowlingApprox} has the well-known Sturm-Liouville form, \eqref{2ndOrderSystem} does not exhibit any particular form. However, in appendix \ref{section:Appendix_Derivation_CSL_Form} we show that making the change of variable
\begin{equation}
\chi \equiv \sqrt{\frac{\eta}{4 \pi G}} \left[\frac{\mathcal{G}'}{\omega_\star^2}+\left(\frac{\omega_\star^{2'}}{\omega_\star^{2}}-k_\star\right) \frac{\mathcal{G}}{\omega_\star^{2}}+\xi\right],
\label{def:chi}
\end{equation}
yields, from \eqref{2ndOrderSystem}, the following Coupled Sturm-Liouville system
\begin{subequations}
	\begin{empheq}{align}
     \left(P_1 \xi'\right)'+ Q_1 \ \! \xi = \sqrt{\frac{\eta}{4 \pi G}} \omega_\star^{4} \ \chi, &
        \label{CSL_a}\\
     \left(P_2 \chi'\right)'+Q_2 \ \! \chi = \sqrt{\frac{\eta}{4 \pi G}} \omega_\star^{4} \ \xi, &
        \label{CSL_b}
    \end{empheq}
\label{CSL}
\end{subequations}
where
\begin{equation}
\renewcommand\arraystretch{1.8} 
\begin{array}{l}
\displaystyle P_1 = P,\\
\displaystyle Q_1 = Q + \frac{\eta}{4 \pi G} \omega_\star^{4}, \\
\end{array}
\label{CSL_Coeffs_1}
\end{equation}
and
\begin{equation}
\renewcommand\arraystretch{1.8} 
\begin{array}{l}
\displaystyle P_2 = \frac{\omega_\star^{4}}{2 \frac{\omega_\star^{2'}}{\omega_\star^{2}} \left(\frac{\omega_\star^{2'}}{\omega_\star^{2}} - k_\star\right) - \frac{\omega_\star^{2''}}{\omega_\star^{2}} + k_\star^2 + k_\star' - k_0^2 \left(1+\eta \ \! \frac{\omega_\star^{2}}{\omega^2}\right)},\\
\displaystyle Q_2 = (k_\star P_2)' - k_\star^2 P_2 + \omega_\star^{4}. \\
\end{array}
\label{CSL_Coeffs_2}
\end{equation}
This is a rigorous mathematical result, and also represents an important improvement on previous derivations on self-gravitating, magnetized plane-parallel configurations. The first line of \eqref{CSL} corresponds in essence to the momentum equation and the second line to Poisson's equation. Interestingly, the form \eqref{CSL} is symmetric, in the sense that a single coupling parameter enters both equations, namely $\sqrt{\eta/4 \pi G} \ \omega_\star^{4}$.
The fact that this parameter is proportional to $\omega_0^2$ is natural, because it means that in systems such as stars and laboratory plasmas in which the free-fall timescale is long, i.e. $\omega_0^2$ is small, the coupling is small. In this case we thus have that while $\eta=1$ (i.e. without the Cowling approximation) the coupling $\sqrt{\eta/4 \pi G} \ \omega_\star^{4}  \rightarrow 0$, but this behavior is reproduced if we simply take $\eta \rightarrow 0$, i.e. if we make the Cowling approximation. In other words, the dependency in $\omega_0^2$ of the coupling term in \eqref{CSL} is consistent with the fact that the Cowling approximation does satisfactorily approximate the evolution of oscillations in systems such as stars and laboratory plasmas.

Having obtained \eqref{CSL}, a natural next step would be to take advantage of the literature on systems of differential equations and higher order Sturm-Liouville problems. \cite{Chuan82,Chuan88,Chuan92} could be a starting point, as well as \cite{Pryce94} on vector Sturm-Liouville problems. Another mathematical method that could be useful is the classical Infeld-Hull factorization method \citep{InfeldHull51} generalized by \cite{Humi86} to coupled systems of second-order differential equations. Analyzing \eqref{CSL} by means of ladder (creation/annihilation) operators seems natural given the formal analogy with quantum mechanics, in which coupled Sturm-Liouville forms arise frequently. For instance \cite{LandauLifshitz81} obtain a set of equations similar to \eqref{CSL}, yet much simpler, when analyzing a hydrogen atom in an electric field. In the same spirit, a formal analogy with quantum mechanical harmonic oscillators will be pointed out in the next section.

Another essential feature to scrutinize are the possible existence of singularities of the coupled ordinary differential equation form \eqref{CSL}. Expressions \eqref{CSL_Coeffs_2} are a compact way of writing these coefficients down, but in order to identify the singularities of this system we should, as done for $P_1$, write $P_2$ as a ratio of polynomials in $\omega^2$. Making explicit $\omega_\star^{2}$ and $k_\star$ from their definitions \eqref{def:omegastar_and_kstar_Explicit} we get that it is a ratio of two polynomials of order 8 in $\omega^2$, namely
\begin{equation}
P_2 = \frac{N_2}{D_2},
\end{equation}
where the numerator contains only the Alfv\'en singularity and simply reads
\begin{equation}
N_2 = \omega^8 \omega_0^8 (\omega^2-\omega_A^2)^4,
\end{equation}
while the denominator is far more involved, as it may be written as
\begin{equation}
D_2 = D \ p_6(\omega^2),
\end{equation}
where the factor $D$ from \eqref{Def:D} returns, and $p_6$ is a sixth order polynomial with lengthy coefficients, which are straightforward to obtain from \eqref{CSL_Coeffs_2} but that are not worth showing explicitly for our purpose here.

Let us review the number of waves that we expect. Compared to the Cowling case, the number of $\partial_t$ terms in the governing partial differential equations is the same, so the number of waves also. In fact, we have seven time derivatives, and one constraint due to $\vec{\nabla} \cdot \vec{B} = 0$. Hence there are six waves. We eliminated the entropy one when stating relation \eqref{ClosureRelationNoBuoyancy}. The Poisson equation does not add a constraint in that sense, and does not add waves because it is the elliptic limit from the hyperbolic General relativistic case.

Let us now investigate what the highest order term of the wave equation should look like. From \eqref{CSL} it is straigthforward to derive the final fourth order scalar equation\footnote{See section \ref{sec:WaveEquation} for a full expression of the scalar wave equation.}, by simply injecting the expression of $\chi$ from \eqref{CSL_a} into \eqref{CSL_b}. Doing so, it immediately appears that the coefficient of the highest order term is proportional to $P_1 P_2/\omega_\star^{4}$, which is itself proportional to (i.e. discarding quantities independent of the eigenvalue $\omega^2$)
\begin{equation}
\frac{\omega^4 (\omega^2-\omega_A^2)^3 (\omega^2-\omega_S^2)}{p_6(\omega^2)}.
\label{CoeffHighestOrderTerm}
\end{equation}
Thus, the slow and Alfv\'en singularities appear in the numerator of the highest order term, as they did in the second order differential equation \eqref{WaveEquation_CowlingApprox} of the Cowling case. This strongly suggests that the slow and Alfv\'en singularities remain intact as genuine continua, as in the Cowling case. In other words, they are unaltered by the perturbation of the gravitational field, as claimed in \cite{PoedtsEtAl85}.

Now, let us focus on the denominator of \eqref{CoeffHighestOrderTerm}. In the Cowling case \eqref{WaveEquation_CowlingApprox}, $D$ was in the denominator of the highest order term, so by analogy, it seems that the polynomial $p_6(\omega^2)$ plays the role of $D$ once the Jeans term is added. It is thus likely that the roots of $p_6(\omega^2)$ are only apparent. Studying the discriminant of this polynomial, it appears that often some, if not all, of its roots are not real. And since we know that $\omega^2$ is real, by self-adjointess, these roots do not belong to the spectrum.
Finally, for illustration let us consider the hydrodynamical limit. In this case, the polynomial has the very simple form $p_6(X)=(a_1 X + a_0) X^5$ where $a_1=-k_0^2 \rho_0^2$ and $a_0=k_0^2 \rho_0^2 (c^2 k_0^2 -(c^2)''- \eta \omega_0^2)$, such that,
using the hydrodynamical equilibrium \eqref{ForceBalance} which now reads $g_0=c^2 \rho_0'/\rho_0$, we have
\begin{subequations}
	\begin{empheq}{align}
     & \omega_\star^{2}=\frac{\omega^2 \omega_0^2}{\omega^2-\omega_{f0}^2},
        \label{BC_hydro_1_variablesXiandG}\\
     & k_\star = \frac{\rho_0'}{\rho_0},
        \label{BC_hydro_2_variablesXiandG}\\
     & P_1=\frac{\rho_0 c^2 \omega^2}{\omega^2-\omega_{f0}^2},
        \label{BC_hydro_3_variablesXiandG}\\
     & P_2=-\frac{\omega^4 \omega_0^4}{k_0^2 (\omega^2-\omega_{f0}^2)(\omega^2-\omega_G^2)},
        \label{BC_hydro_4_variablesXiandG}
    \end{empheq}
\label{BC_hydro_variablesXiandG}
\end{subequations}
where the fast turning point frequency reduces to $\omega_{f0}^2(x)=c^2(x) k_0^2$, and
\begin{equation}
\omega_G^2(x) \equiv c^2 k_0^2-(c^2)''-\eta \omega_0^2(x).
\end{equation}
As above, we can easily derive the expression for the coefficient of the highest order term of the wave equation satisfied by $\xi$. Up to a factor independent of $\omega^2$ it reads
\begin{equation}
\frac{\omega^2}{\omega^2-\omega_G^2},
\end{equation}
while in the Cowling case, the coefficient of the highest order term of the (second order) wave equation reads, up to a factor independent of $\omega^2$,
\begin{equation}
\frac{\omega^2}{\omega^2-\omega_{f0}^2}.
\end{equation}
From this observation, we are led to conclude that the range of frequencies $\{\omega_G^2\}$ generalizes $\{\omega_{f0}^2\}$ of the Cowling case. In fact, this idea was already suggested in \cite{Durrive17}, who analyzed this hydrodynamical case, except that now we have a magnetized version of this conclusion: from \eqref{CoeffHighestOrderTerm} we conjecture that the roots of $p_6(\omega^2)$ generalize the apparent singularities found in the turning point frequency ranges $\{\omega_{s0}^2\}$ and $\{\omega_{f0}^2\}$ of the Cowling case.

\section{Coupled Harmonic Oscillator form}
\label{sec:CHO}

As a mathematically equivalent form, we can also transfrom the coupled Sturm-Liouville set into a set of equations expressing two coupled harmonic oscillators.
Putting
\begin{equation}
\renewcommand\arraystretch{2.} 
\begin{array}{l}
\displaystyle
\chi_1 \equiv \sqrt{|P_1|} \ \! \xi,\\
\displaystyle
\chi_2 \equiv \sqrt{|P_2|} \ \! \chi,
\end{array}
\label{CHO_def_chi1_chi2}
\end{equation}
we may rewrite \eqref{CSL} into the coupled harmonic oscillator form
\begin{equation}
\renewcommand\arraystretch{2.} 
\begin{array}{l}
\displaystyle
\chi_1''-\kappa_1^2 \ \! \chi_1 = s_1 \ \! \kappa_c^2 \ \! \chi_2,\\
\displaystyle
\chi_2''-\kappa_2^2 \ \! \chi_2 = s_2 \ \! \kappa_c^2 \ \! \chi_1,
\end{array}
\label{CHO}
\end{equation}
where $\kappa_1, \kappa_2$ and $\kappa_c$ are wavenumbers, with the following expressions: the coupling parameter now reads
\begin{equation}
\kappa_c^2 \equiv \sqrt{\frac{\eta}{4 \pi G |P_1 P_2|}} \ \omega_\star^{4},
\label{def:kappa_c}
\end{equation}
and for $i=1,2$
\begin{equation}
\renewcommand\arraystretch{2.3} 
\begin{array}{rl}
\displaystyle
\kappa_i^2 & \equiv \frac{\left(\sqrt{|P_i|}\right)''}{\sqrt{|P_i|}} - \frac{Q_i}{P_i},\\
\displaystyle
 & = \frac{1}{2} \frac{P_i''}{P_i} -\frac{1}{4} \left(\frac{P_i'}{P_i}\right)^2 - \frac{Q_i}{P_i},\\
s_i & \equiv \text{sign}(P_i) \hspace{0.3cm} \text{on the considered interval}.
\end{array}
\label{def:kappa_i}
\end{equation}
The first expression for $\kappa_i^2$ corresponds to the way it naturally appears when performing the calculation, while the second expression is more convenient as it does not contain square roots and absolute values.
Note that the singularities that became evident in the Sturm-Liouville form now appear as frequency ranges where the coupling parameter becomes locally infinite, as seen from \eqref{def:kappa_c}. This is yet another indication that they are physically significant in the MHD spectrum.
The advantages of this reformulation are the following.

Coupled harmonic oscillators often arise in physics, such that \eqref{CHO} may be used to interpret the dynamics of the system through formal analogies with other physical systems (the simplest example being two masses coupled by springs) which could help build an intuition of the behavior of the system. Moreover, these analogies could also be useful technically speaking, as numerous studies are dedicated to finding the solutions of time-dependent\footnote{Which is identical to \eqref{CHO}, replacing $x$ by the time variable.} coupled harmonic oscillators, particularly in quantum physics. To adapt our system to these studies, one should express \eqref{CHO} in Hamiltonian form, as follows.
Using the spatial $x$ coordinate as time, and the rescaled eigenfunctions $\chi_i$'s as generalized coordinates, i.e. defining some abstract canonical coordinates $(q_i,p_i)$ with $q_i = \chi_i$, the system \eqref{CHO} corresponds to Hamilton's equations
\begin{equation}
\renewcommand\arraystretch{2.} 
\begin{array}{l}
\displaystyle
\chi'_i = \frac{\partial \mathcal{H}}{\partial p_i},\\
\displaystyle
p'_i = -\frac{\partial \mathcal{H}}{\partial \chi_i},
\end{array}
\end{equation}
for the Hamiltonian
\begin{equation}
\mathcal{H} = \frac{1}{2} \left[s_2 p_1^2 + s_1 p_2^2 - s_2 \kappa_1^2 \chi_1^2 - s_1 \kappa_2^2 \chi_2^2\right] - s_1 s_2 \kappa_c^2 \chi_1 \chi_2.
\end{equation}
This is indeed the Hamiltonian of two-coupled harmonic oscillators, with coupling $- s_1 s_2 \kappa_c^2$, but with the exotic feature that the kinetic terms can have negative signs. The analysis could then be carried on in the line of \citet[][]{BruschiEtAl19,UrzuaEtAl19,MoyaCessaRecamier20,RamosPrietoEtAl20} for example, to quote only recent studies.

In addition, the coupled harmonic oscillator form may be convenient to derive oscillation theorems regarding the present eigenvalue problem. For example, with $\vec{u} \equiv (\chi_1,\chi_2)^{\textsc{T}}$, expression \eqref{CHO} may be rearranged as
\begin{equation}
\vec{u}'' - A \vec{u} =0,
\label{Upp_Au}
\end{equation}
where $A$ is the real matrix
\begin{equation}
A =
\left( \!
\renewcommand\arraystretch{1.4} 
\begin{array}{cc}
\kappa_1^2 & s_1 \ \! \kappa_c^2 \\
s_2 \ \! \kappa_c^2 & \kappa_2^2 \\
\end{array}
\! \right),
\label{matrixA}
\end{equation}
which corresponds to the type of differential systems analyzed by \cite{KeenerTravis80} for instance. This paper notably illustrates the fact that the Sturmian properties of an equation like \eqref{Upp_Au} depend on whether $A$ is symmetric or not. From this, we note the following intriguing feature. On the one hand, the symmetry of $A$ is directly given by the signs $s_1$ and $s_2$, as seen in definition \eqref{matrixA}. On the other hand, in the Cowling approximation, it was shown that the Sturmian properties of the wave equation are directly related to the sign $s_1$ (i.e. to the sign of $P$ since $P_1=P$), as discussed below equation \eqref{WaveEquation_CowlingApprox}. Therefore, the aforementioned result from \cite{KeenerTravis80} 
resembles the foundations of a yet-to-be-constructed oscillation theorem generalizing that of \cite{GoedbloedSakanaka74} to the present differential system. We will not explore this further here, but it indicates how reformulations may indeed help obtaining a complete rigorous derivation of the spectrum of the force operator~\eqref{ForceOperator}.

\section{Matrix differential equation}
\label{sec:MatrixDifferentialEquation}

In the initial formulation \eqref{MatrixOperatorForm}, the $4 \times 4$ matrix involved was a matrix operator, of second order. Thanks to the above reformulations, we are now in a position to reformulate the problem by means of an algebraic $4 \times 4$ matrix instead, which is a significant simplification. Indeed, defining the vector
\begin{equation}
\vec{V} \equiv (\chi_1',\chi_2',\chi_1,\chi_2)^{\textsc{T}},
\end{equation}
the system \eqref{CHO} can be rewritten in the first order $4 \times 4$ matrix form
\begin{equation}
\vec{V}'
=
\mathsf{M}_\textsc{v} \vec{V},
\label{CHO_4times4_MatrixForm}
\end{equation}
with the simple block matrix
\begin{equation}
\mathsf{M}_\textsc{v} = \left( \!
\renewcommand\arraystretch{1.4} 
\begin{array}{cc}
0 & A \\
1 \! \! 1 & 0 
\end{array}
\! \right),
\label{Matrix_MV}
\end{equation}
where $1 \! \! 1$ is the $2 \times 2$ identity matrix, and $A$ is given by \eqref{matrixA}. The important difference with \eqref{MatrixOperatorForm} is that now we may obtain the solutions explicitly by means of standard expansions. Examples of general solutions are given in appendix~\ref{appendix:ExplicitingSolutions}, but for illustration, let us here show the solution explicitly in a particular regime, namely when the typical lengthscale $L$ of variation of the coefficients in matrix $A$ is large compared to the thickness of the slab, i.e. $x_b \ll L$ where $x_b$ denotes the position of the boundary (the slab lies in $[-x_b,x_b]$). In this case, the slab is thin enough for the matrix $\mathsf{M}_\textsc{v}$ in \eqref{CHO_4times4_MatrixForm} to be roughly constant, i.e. $\mathsf{M}_\textsc{v}(x) \approx \mathsf{M}_\textsc{v}(0)$ throughout the slab, and the solution is well approximated by
\begin{equation}
\vec{V}(x)=e^{x \mathsf{M}_\textsc{v}(0)} \vec{V}(0).
\label{V_ThinLimit_MatrixForm}
\end{equation}
Physically, we expect this regime to be relevant for slabs subject to a strong external pressure, e.g. in a cloud-cloud collision, in a supernovae remnant, or under strong radiation pressure. Expression \eqref{V_ThinLimit_MatrixForm} is compact, but a priori it is not easy to manipulate, since it requires computing the exponential of a $4 \times 4$ matrix. Fortunately however, because the matrix $\mathsf{M}_\textsc{v}$, given by \eqref{Matrix_MV}, has a simple form, the solution \eqref{V_ThinLimit_MatrixForm} may be made explicit further as follows. The eigenvalues of $\mathsf{M}_\textsc{v}$ are
\begin{equation}
\left( \! \!
\begin{array}{c}
\mathbb{K}_1\\
\mathbb{K}_2\\
\mathbb{K}_3\\
\mathbb{K}_4
\end{array}
\! \! \right)
\equiv
\left( \! \!
\begin{array}{r}
-\kappa_{\m}\\
\kappa_{\m}\\
-\kappa_{\p}\\
\kappa_{\p}
\end{array}
\! \! \right),
\end{equation}
where
\begin{equation}
\kappa_{\pm} \equiv \sqrt{\frac{\kappa_1^2+\kappa_2^2 \pm \kappa_\delta^2}{2}},
\label{def:kappa_pm}
\end{equation}
and
\begin{equation}
\kappa_\delta^2 \equiv \sqrt{(\kappa_1^2-\kappa_2^2)^2+4 s_1 s_2 \kappa_c^4}.
\end{equation}
We introduced the symbols $\kappa$ and $\mathbb{K}$, because these symbols are close to the letter $k$, reminding that they have the dimension of wavenumbers. The corresponding eigenvectors are
\begin{equation}
\vec{\sigma}_1 =
\left( \! \!
\begin{array}{c}
-\kappa_{\m} \Delta_{\m}\\
-\kappa_{\m} \\
\Delta_{\m}\\
1
\end{array}
\! \! \right)
\text{,} \hspace{0.5cm}
\vec{\sigma}_2 =
\left( \! \!
\begin{array}{c}
\kappa_{\m} \Delta_{\m}\\
\kappa_{\m} \\
\Delta_{\m}\\
1
\end{array}
\! \! \right),
\end{equation}
and
\begin{equation}
\vec{\sigma}_3 =
\left( \! \!
\begin{array}{c}
-\kappa_{\p} \Delta_{\p}\\
-\kappa_{\p} \\
\Delta_{\p}\\
1
\end{array}
\! \! \right)
\text{,} \hspace{0.5cm}
\vec{\sigma}_4 =
\left( \! \!
\begin{array}{c}
\kappa_{\p} \Delta_{\p}\\
\kappa_{\p} \\
\Delta_{\p}\\
1
\end{array}
\! \! \right),
\end{equation}
with the dimensionless parameter
\begin{equation}
\displaystyle \Delta_{\pm} \equiv \frac{2 s_1 \ \! \kappa_c^2}{\kappa_2^2-\kappa_1^2 \pm \kappa_\delta^2}.
\end{equation}

\noindent
The diagonalization of $\mathsf{M}_\textsc{v}$ reads
\begin{equation}
\mathsf{M}_\textsc{v}(0)=\vec{P} \vec{D} \vec{P}^{-1},
\label{diagonalization}
\end{equation}
where $\vec{D}~\equiv~\text{diag}(-\kappa_{\m},\kappa_{\m},-\kappa_{\p},\kappa_{\p})$ and $\vec{P}$ is the matrix whose columns are the eigenvectors $\vec{\sigma}_i$'s. Then, with the shorthand notation
\begin{equation}
c_{\pm} \equiv \cosh(\kappa_{\pm} x) \hspace{0.3cm} \text{ and } \hspace{0.3cm} s_{\pm} \equiv \sinh(\kappa_{\pm} x),
\end{equation}
we have that \eqref{V_ThinLimit_MatrixForm} may be written as

\begin{widetext}
\begin{equation}
\renewcommand\arraystretch{1.5} 
\vec{V}(x)=
\frac{1}{\Delta_{\m} \! \! - \! \! \Delta_{\p}} \! \! 
\left( \! \!
\begin{array}{rrrr}
\Delta_{\m} c_{\m}-\Delta_{\p} c_{\p} & \Delta_{\p} \Delta_{\m} c_{\p}- \Delta_{\m} \Delta_{\p} c_{\m} & \Delta_{\m} \kappa_{\m} s_{\m}-\Delta_{\p} \kappa_{\p} s_{\p} & \Delta_{\p} \Delta_{\m}  \kappa_{\p} s_{\p}- \Delta_{\m} \Delta_{\p} \kappa_{\m} s_{\m}\\
\hspace{0.48cm} c_{\m}- \hspace{0.48cm} c_{\p} & \hspace{0.48cm} \Delta_{\m} c_{\p}- \hspace{0.48cm} \Delta_{\p} c_{\m} & \hspace{0.48cm} \kappa_{\m} s_{\m}- \hspace{0.48cm} \kappa_{\p} s_{\p} & \hspace{0.48cm} \Delta_{\m} \kappa_{\p} s_{\p} - \hspace{0.48cm} \Delta_{\p} \kappa_{\m} s_{\m}\\
\Delta_{\m} \frac{s_{\m}}{\kappa_{\m}}-\Delta_{\p} \frac{s_{\p}}{\kappa_{\p}} & \Delta_{\p} \Delta_{\m} \frac{s_{\p}}{\kappa_{\p}}- \Delta_{\m} \Delta_{\p} \frac{s_{\m}}{\kappa_{\m}} & \Delta_{\m} c_{\m}-\Delta_{\p} c_{\p} & \Delta_{\p} \Delta_{\m} c_{\p}- \Delta_{\m} \Delta_{\p} c_{\m}\\
\hspace{0.48cm} \frac{s_{\m}}{\kappa_{\m}}- \hspace{0.48cm} \frac{s_{\p}}{\kappa_{\p}} & \hspace{0.48cm} \Delta_{\m} \frac{s_{\p}}{\kappa_{\p}}- \hspace{0.48cm} \Delta_{\p} \frac{s_{\m}}{\kappa_{\m}} & \hspace{0.48cm} c_{\m}- \hspace{0.48cm} c_{\p} & \hspace{0.48cm} \Delta_{\m} c_{\p}- \hspace{0.48cm} \Delta_{\p} c_{\m}
\end{array}
\! \! \right) \vec{V}(0).
\label{V_ThinLimit_MatrixForm_Version1}
\end{equation}
\end{widetext}
This is a lenghtier expression than \eqref{V_ThinLimit_MatrixForm} but it is significantly simpler to manipulate.
Note that since $\mathsf{M}_\textsc{v}$ is evaluated at $x=0$ in \eqref{V_ThinLimit_MatrixForm}, the $x$-dependence in \eqref{V_ThinLimit_MatrixForm_Version1} lies only in $c_{\pm}$ and $s_{\pm}$.
The expression \eqref{V_ThinLimit_MatrixForm_Version1} is an important result since from it we can derive fully analytic solutions to the eigenvalue-eigenfunction problem of a self-gravitating, thin, magnetized slab in all its generality. Now, because the slab is thin, the complexities of the continuous ranges are in essence avoided, and the focus is instead on the discrete modes (which may obey anti-sturmian properties in certain frequency regimes). We exemplify this in the hydrodynamical case in section \ref{sec:Example}.

\section{Scalar wave equation}
\label{sec:WaveEquation}

As a last way to mathematically reformulate our problem, we can state the final fourth-order scalar wave equation, directly generalizing the scalar wave equation number (7.80) in \cite{GKP}. In \eqref{CHO}, injecting the expression of $\chi_2$ from the first equation into the second equation gives
\begin{equation}
\left(\alpha_4 \chi_1''\right)''+\alpha_2 \ \! \chi_1'' + \alpha_1 \ \! \chi_1' + \alpha_0 \ \! \chi_1 = 0
\label{WaveEquation_FromCHO}
\end{equation}
where
\begin{equation}
\renewcommand\arraystretch{1.3} 
\begin{array}{l}
\displaystyle
\alpha_4 \equiv \kappa_c^{-2},\\
\displaystyle
\alpha_2 \equiv -\kappa_c^{-2} (\kappa_1^2+\kappa_2^2),\\
\displaystyle
\alpha_1 \equiv -2 (\kappa_c^{-2} \kappa_1^2)',\\
\displaystyle
\alpha_0 \equiv \kappa_1^2 \kappa_2^2 \kappa_c^{-2} -(\kappa_c^{-2} \kappa_1^2)'' - s_1 s_2 \kappa_c^{2}.
\end{array}
\end{equation}
This equation should be seen as an equation on $\xi$, using  $\chi_1 = \sqrt{|P_1|} \ \! \xi$ from the definition~\eqref{CHO_def_chi1_chi2}. 
The strength of the expression \eqref{WaveEquation_FromCHO} is its compactness. Indeed, in order to appreciate how lengthy this full-gravity MHD wave equation is, we invite the reader to collect the definitions of the quantities inside the coefficients $\alpha_i$
(using relations \eqref{def:omegastar_and_kstar_Explicit}, \eqref{CSL_Coeffs_1}, \eqref{CSL_Coeffs_2}, \eqref{def:kappa_c}, \eqref{def:kappa_i}), and expand all the derivatives, to figure out this equation explicitly in terms of the original equilibrium quantities $\rho_0(x),c(x),b(x),B_0(x),g_0(x)$, of the wavenumbers $k_\perp(x),k_\parallel(x)$, and of the eigenvalue $\omega^2$. This should highlight the necessity of having gone through all the above reformulations.
Finally, analyzing expression \eqref{WaveEquation_FromCHO} further may be done in the light of the literature on non-selfadjoint fourth order differential equations such as for instance \citet{Kreith74_a,Kreith74_b,ChengEdelson78,KeenerTravis80}.

\section{Solutions in terms of the Lagrangrian displacements}
\label{sec:FinalStep}

We reformulated the problem of all eigenoscillations of a self-gravitating magnetized slab into various compact, classical forms, notably by means of the $(\chi_1,\chi_2)$ variables of equations \eqref{CHO_def_chi1_chi2}.
In so doing, we were able to obtain an explicit analytic form in a special (thin slab) case, namely \eqref{V_ThinLimit_MatrixForm_Version1}.
Let us now assume that we have such an explicit expression for the vector $\vec{V} = (\chi_1',\chi_2',\chi_1,\chi_2)^{\textsc{T}}$. The final step is then to express the initial variables $(\xi,\xi_\perp,\xi_\parallel)$ and $(g_x,g_\perp,g_\parallel)$ (defined in \eqref{def:xi_components} and \eqref{def:g_components}) that we are ultimately looking for, in terms of $(\chi_1',\chi_2',\chi_1,\chi_2)$.

Using \eqref{XiPerp_XiParallel} we have the link between $(\xi_\perp,\xi_\parallel)$ and $(\mathcal{G},\xi',\xi)$, which may be made explicit looking at the components of \eqref{MatrixF}.  Then in appendix \ref{appendix:BackToInitialVariables} we show the link between $(\mathcal{G}',\mathcal{G},\xi',\xi)$ and $(\chi_1',\chi_2',\chi_1,\chi_2)$. Putting this result into a single matrix product form, we get the final expression
\begin{equation}
\begin{array}{l}
\left(
\begin{array}{c}
\xi\\
\xi_\perp \\
\xi_\parallel
\end{array}
\right)
= \mathsf{M}_\xi \vec{V},
\end{array}
\label{def:xi_components_FINALRESULT}
\end{equation}
and relations \eqref{FieldLineProjection_of_irrotationalConstraint} and \eqref{gperp_gpara} in matrix form read
\begin{equation}
\begin{array}{l}
\left(
\begin{array}{c}
g_x\\
g_\perp \\
g_\parallel
\end{array}
\right)
= \mathsf{M}_g \vec{V},
\end{array}
\label{def:g_components_FINALRESULT}
\end{equation}
where the transformation matrices $\mathsf{M}_\xi$ and $\mathsf{M}_g$ are given by
\begin{equation}
\mathsf{M}_\xi = \mathsf{M}_1 \mathsf{M}_2 \mathsf{M}_3,
\end{equation}
where
\begin{equation}
\mathsf{M}_1
=
\left(
\arraycolsep=1.4pt 
\begin{array}{cccc}
1 & 0 & 0 \\
0 & (\omega^2 - k_\parallel^2 c^2)/D & k_\parallel k_\perp c^2/D \\
0 & k_\parallel k_\perp c^2/D & (\omega^2 - k_\perp^2 (c^2 \! + \! b^2) - k_\parallel^2b^2)/D \\
\end{array}
\right),
\label{def:matrix_M1}
\end{equation}
and
\begin{equation}
\mathsf{M}_2
=
\left(
\begin{array}{cccc}
0 & 0 & 0 & 1 \\
0 & -\eta k_\perp & k_\perp (c^2+b^2) & k_\perp g_0 \\
0 & -\eta k_\parallel & k_\parallel c^2 & k_\parallel g_0
\end{array}
\right),
\label{def:matrix_M2}
\end{equation}
and
\begin{equation}
\mathsf{M}_3
=
\left(
\begin{array}{cccc}
0 & \tau_{12} & \tau_{13} & \tau_{14} \\
0 & \tau_{22} & 0 & \tau_{24} \\
\tau_{31} & 0 & \tau_{33} & 0 \\
0 & 0 & \tau_{43} & 0
\end{array}
\right),
\label{def:matrix_M3}
\end{equation}
and
\begin{equation}
\mathsf{M}_g
=
\left(
\begin{array}{crcr}
0 & -\tau_{12} & -\tau_{13} & -\tau_{14} \\
0 & k_\perp \tau_{22} & 0 & k_\perp \tau_{24} \\
0 & k_\parallel \tau_{22} & 0 & k_\parallel \tau_{24}
\end{array}
\right),
\label{def:matrix_Mg}
\end{equation}
where the $\tau_{ij}$ coefficients are functions of the parameters of the problem, namely
\begin{equation}
\renewcommand\arraystretch{2.3} 
\begin{array}{l}
\displaystyle
\tau_{12}= \sqrt{\frac{\eta}{4 \pi G}} \left(k_\star-\frac{\omega_\star^{2'}}{\omega_\star^{2}}\right) \tau_{22},\\
\displaystyle
\tau_{13}=-\omega_\star^{2} \ \! \tau_{31},\\
\displaystyle
\displaystyle
\tau_{14}= \frac{\omega_\star^{2}}{\sqrt{|P_2|}} \left[\sqrt{\tfrac{4 \pi G}{\eta}}- s_2 \ \! \tau_{12} \ \! \tau_{24}\right],\\
\displaystyle
\tau_{22}=-\sqrt{\frac{4 \pi G}{\eta}} \frac{\sqrt{|P_2|}}{s_2 \ \! \omega_\star^{2}},\\
\displaystyle
\tau_{24}= \left(k_\star-\frac{P_1'}{2 P_1}\right) \tau_{22},\\
\displaystyle
\tau_{31}=\tau_{43}= \frac{1}{\sqrt{|P_1|}},\\
\displaystyle
\tau_{33}=\tau_{31}'.
\end{array}
\end{equation}
We have thus managed to express the sought Lagrangian displacement vector $\vec{\xi}$ and the perturbed gravitational acceleration $\vec{g}_1$ in terms of known quantities.

\section{A first hydrodynamic example}
\label{sec:Example}

So far, we presented several mathematically equivalent, physically useful forms for the differential system governing the eigenvalue problem. We now illustrate how these help solving the full eigenvalue problem, i.e. taking boundary conditions into account and yielding the dispersion relation. In this section our purpose is (i) to illustrate how Jeans' instability emerges in our formalism, and (ii) to check that we do recover some well-known results from the literature.
As a proof of concept we choose an isothermal thin slab without magnetic field under so-called rigid boundary conditions. 

Limiting the problem to the isothermal hydrodynamical case is convenient because it avoids the complexity of having to deal with the continuous spectra (slow and alfven) altogether. This is obviously why this case has been the one featuring in many previous studies. Also, the equilibrium state is then fully analytical. For instance the self-gravitating density field reads
\begin{equation}
\rho(x)=\rho_c \cosh^{-2} \left(\frac{x \ k_\textsc{J}}{\sqrt{2}}\right),
\label{IsothermalEquilibriumDensity}
\end{equation}
where, introducing
\begin{equation}
\omega_c^2 \equiv \omega_0^2(x=0),
\end{equation}
the Jeans wavenumber
\begin{equation}
k_\textsc{J}^2 \equiv \frac{\omega_c^2}{c^2},
\end{equation}
is defined at the center of the slab.

\subsection{Symmetry and boundary conditions}

The governing differential equation being of order four, we need to impose four conditions. In this section we describe boundary and symmetry conditions commonly used in the literature \citep[see e.g.][for details]{GoldreichLyndenBell65,ElmegreenElmegreen78,KimEtAl12,DurriveLanger19}.
We will in particular compare our analytical results to those of \cite{KimEtAl12} who analyze the Jeans instability in unmagnetized rotating pressure-confined polytropic gas disks. They solve the eigenvalue problem numerically but they also derive approximate analytical results.

As in \cite{KimEtAl12} we restrict ourselves to symmetric modes, which translates into $\hat{\xi}_x(0)=0$ and $\hat{g}_{1x}(0)=0$, i.e. basically the center of the slab is fixed. Considering rigid boundary conditions, we also prevent the surface at the boundary to move, i.e. $\hat{\xi}_x(x_b)=0$, where $x_b$ is the position of the boundary. The fourth condition is more subtle. To obtain it, one needs to apply the divergence theorem to the linearized Poisson equation for an infinitesimally thin shell containing the boundary layer, and compute the gravitational acceleration outside the slab by solving Laplace's equation, using the fact that it should not diverge at infinity and assuming the external fluid remains unperturbed. This results in the constraint $\hat{g}_{1x}(x_b)-i \hat{g}_{1y}(x_b)=0$. There is no $\hat{g}_{1z}$ because in the hydrodynamical limit directions $y$ and $z$ are undistinguishable such that we can rotate the coordinates to remove any dynamics in the $z$-direction without loss of generality and use a normal mode decomposition with $k_y = k_0$ and $k_z = 0$.

For the isothermal hydrodynamical fluid under consideration, in terms of our variables $\xi$ and $\mathcal{G}$ these four conditions become, in the same order as presented above,
\begin{subequations}
	\begin{empheq}{align}
     & \xi(0) = 0,
        \label{BC_hydro_1_variablesXiandG}\\
     & \mathcal{G}'(0)=0,
        \label{BC_hydro_2_variablesXiandG}\\
     & \xi(x_b) = 0,
        \label{BC_hydro_3_variablesXiandG}\\
     & \mathcal{G}'(x_b)+k_0 \mathcal{G}(x_b)=0,
        \label{BC_hydro_4_variablesXiandG}
    \end{empheq}
\label{BC_hydro_variablesXiandG}
\end{subequations}
where we have used relations \eqref{def:xi_components}, \eqref{def:g_components}, \eqref{FieldLineProjection_of_irrotationalConstraint}, and \eqref{def:variableG}.
Then, with definitions \eqref{def:chi} and \eqref{CHO_def_chi1_chi2}, this translates into
\begin{subequations}
	\begin{empheq}{align}
     & \chi_1(0) = 0,
        \label{BC_hydro_isothermal_1}\\
     & \chi_2(0) = 0,
        \label{BC_hydro_isothermal_2}\\
     & \chi_1(x_b) = 0, 
        \label{BC_hydro_isothermal_3}\\
     & \chi'_2(x_b) - \kappa_b \ \! \chi_2(x_b) = 0,
        \label{BC_hydro_isothermal_4}
    \end{empheq}
\label{BC_hydro_isothermal}
\end{subequations}
where a wavenumber, evaluated at $x_b$, appears
\begin{equation}
\left. \kappa_b \equiv -\frac{1}{2} \frac{\rho'_0}{\rho_0} - k_0\left(1+\eta \frac{\omega_0^2}{\omega^2-k_0^2 c^2}\right)\right|_{x=x_b}.
\label{def:kb}
\end{equation}

\hspace{1cm}

\subsection{Dispersion relation}

To derive the dispersion relation, let us use the solution \eqref{V_ThinLimit_MatrixForm}. Using the diagonalization \eqref{diagonalization}, it also reads
\begin{equation}
\vec{V}(x)=\sum_{i=1}^4 \beta_i e^{\mathbb{K}_i x} \vec{\sigma}_i,
\label{V_ThinLimit_bis}
\end{equation}
where $\mathbb{K}_i$ and $\vec{\sigma}_i$ are the eigenelements of the matrix $\mathsf{M}_\textsc{v}$, shown explicitly in section~\ref{sec:MatrixDifferentialEquation}, and the $\beta_i$'s are constants, constrained by the boundary conditions.
With \eqref{V_ThinLimit_bis}, boundary conditions \eqref{BC_hydro_isothermal} can be rearranged as
\begin{equation}
\vec{M}_\textsc{BC} \cdot \vec{\beta} = \vec{0},
\end{equation}
where $\vec{\beta} \equiv (\beta_1,\beta_2,\beta_3,\beta_4)^\textsc{T}$ and, most importantly,

\begin{widetext}
\begin{equation}
\vec{M}_\textsc{BC}(k_0,\omega^2)
\equiv
\left( \! \!
\begin{array}{cccc}
-(\kappa_{\m}+\kappa_b) e^{-x_b \kappa_{\m}} & (\kappa_{\m}-\kappa_b) e^{x_b \kappa_{\m}} & -(\kappa_{\p}+\kappa_b) e^{-x_b \kappa_{\p}} & (\kappa_{\p}-\kappa_b) e^{x_b \kappa_{\p}}\\
\Delta_{\m} e^{-x_b \kappa_{\m}} & \Delta_{\m} e^{x_b \kappa_{\m}} & \Delta_{\p} e^{-x_b \kappa_{\p}} & \Delta_{\p} e^{x_b \kappa_{\p}}\\
\Delta_{\m} & \Delta_{\m} & \Delta_{\p} & \Delta_{\p} \\
1 & 1 & 1 & 1
\end{array}
\! \! \right).
\end{equation}
\end{widetext}

\noindent
Pairs of parameters $(k_0,\omega^2)$ for which $\vec{M}_\textsc{BC}$ is not invertible yield non-trivial solutions satisfying the boundary conditions. In other words, the dispersion relation is given by
\begin{equation}
\det(\vec{M}_\textsc{BC}) = 0.
\label{DispRel}
\end{equation}
This
$\vec{M}_\textsc{BC}$ is simple enough for its determinant to be computed by hand.
Doing so relation \eqref{DispRel} gives
\begin{equation}
\renewcommand\arraystretch{1.2} 
\begin{array}{l}
\sinh(x_b \kappa_{\p}) (\kappa_{\p}^2-\kappa_2^2)\left[\kappa_{\m} \cosh(x_b \kappa_{\m})-\kappa_b \sinh(x_b \kappa_{\m})\right]\\
\hspace{-0.75cm} \textcolor{white}{\frac{k_0}{k_0}} - \sinh(x_b \kappa_{\m}) (\kappa_{\m}^2-\kappa_2^2)\left[\kappa_{\p} \cosh(x_b \kappa_{\p})-\kappa_b \sinh(x_b \kappa_{\p})\right]=0.
\end{array}
\label{Disp_rel_thinlimit}
\end{equation}
Thus, we have reduced our eigenvalue problem to simply the determination of the roots of the above left-hand side, seen as a function of $\omega^2$ and parametrized by $k_0$.
In the next two sections we show that in \eqref{Disp_rel_thinlimit} the second term dominates for high frequencies and corresponds to the p-modes, while the first term dominates for low frequencies, and corresponds to gravitational instability.

\subsection{High frequency limit: p-modes}
\label{sec:HF_limit}

We know that \eqref{Disp_rel_thinlimit} must contain an infinite number of solutions, because it must at least contain the p-modes. The key is that the quantities entering this relation are complex numbers, such that the hyperbolic functions $\sinh$ and $\cosh$ become regular sine and cosine functions once their argument is purely imaginary. For instance, in the high frequency regime ($\omega^2 \to \infty$), $\kappa_2$ and $\kappa_b$ are independent of $\omega$, while $\kappa_{\m} \sim i \omega/c$ and $\kappa_{\p}\sim k_0$, so that the second term in  \eqref{Disp_rel_thinlimit} dominates, and we are left with
\begin{equation}
\sin(x_b |\kappa_{\m}|)=0.
\label{D_HF}
\end{equation}
The sine function appears because $\kappa_{\m}$ is purely imaginary in this regime.
The constraint \eqref{D_HF} yields, with $n \in \mathbb{N}$,
\begin{equation}
x_b |\kappa_{\m}| = n \pi.
\end{equation}
Taking the square of this relation and Taylor expanding $\kappa_{\m}$ in $\omega$, we get the high frequency behavior of the eigenvalues
\begin{equation}
\omega^2_n=\left(\frac{n^2 \pi^2}{x_b^2}+k_0^2\right) c^2+\left(\frac{1}{2}- \eta \right) \omega_c^2.
\label{HFspectrum_thin_limit}
\end{equation}
We recognize in this expression the p-modes, with a slight correction (the term with $\eta$) due to the Jeans term.
In figure~\ref{fig:Example} we compare the predicted high-frequency spectrum \eqref{HFspectrum_thin_limit} to the numerical resolution of this eigenvalue problem, and the two results are in agreement.
The result \eqref{HFspectrum_thin_limit} generalizes a classical result in the literature \citep[e.g.][]{KimEtAl12,Durrive17}. Indeed, in the Cowling approximation $\eta=0$, and only then, the present eigenvalue problem (planar, hydrodynamical, isothermal, with rigid boundary conditions) can be solved exactly, even without the small thickness assumption, because luckily the second order differential equation involved has a simple analytical expression. The result of this calculation yields precisely \eqref{HFspectrum_thin_limit} with $\eta=0$. To the best of our knowledge, the present generalization is new.

\subsection{Gravitational instability}

On the contrary, if we now consider a finite $\omega$, given that $x_b$ is small we can reduce \eqref{Disp_rel_thinlimit} to
\begin{equation}
\kappa_b x_b-1=0.
\label{kbxbEqual1}
\end{equation}
Looking at the definition \eqref{def:kb} of $\kappa_b$, it appears that \eqref{kbxbEqual1} is a constraint on $\omega^2$ with only one solution. Let us call this solution $\omega^2_\text{GI}$ because it corresponds to the mode prone to Jeans' gravitational instability. Explicitly we get
\begin{equation}
\omega^2_\text{GI} = k_0^2 c^2 - \eta \frac{4 \pi G \rho_b x_b k_0}{1+ x_b \left(k_0+ \left. \frac{1}{2} \frac{\rho'_0}{\rho_0}\right|_{x=x_b}\right)}.
\end{equation}
Furthermore, using the equilibrium density \eqref{IsothermalEquilibriumDensity} and expanding up to second order in $x_b$ and in $k_0$ (because only small wavenumbers matter for this instability) gives
\begin{equation}
\omega^2_\text{GI} = c^2_\text{eff} k_0^2 + g_\text{eff} \ \! k_0,
\label{omega2GI_thinLimit}
\end{equation}
where
\begin{equation}
\renewcommand\arraystretch{1.2} 
\begin{array}{l}
\displaystyle
c^2_\text{eff} \equiv c^2+ \eta \omega_c^2 x_b^2, \\
\displaystyle
g_\text{eff} \equiv - \eta \omega_c^2 x_b,
\end{array}
\label{omega2GI_thinLimit_Coefficients}
\end{equation}
are an effective speed of sound and an effective gravitational acceleration, respectively.

\begin{figure*}
\centering
\begin{minipage}{.5\textwidth}
  \centering
  \includegraphics[width=0.95\linewidth]{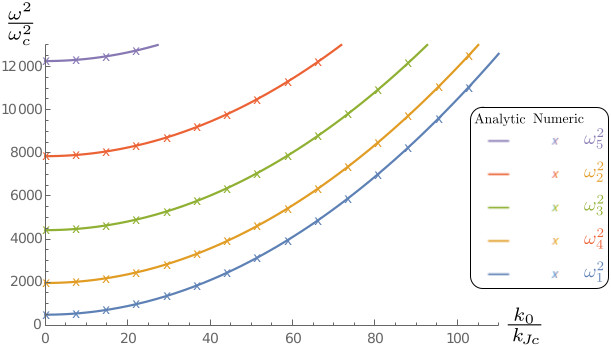}
\end{minipage}%
\begin{minipage}{.5\textwidth}
  \centering
  \includegraphics[width=0.95\linewidth]{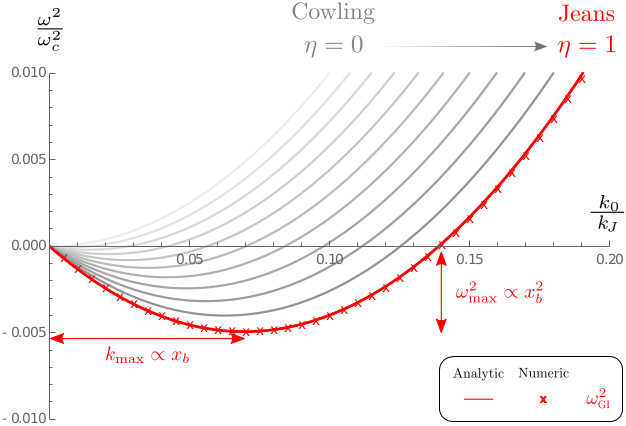}
\end{minipage}
\caption{Plots of the spectrum deduced analytically in the thin limit (expressions \eqref{HFspectrum_thin_limit} and \eqref{omega2GI_thinLimit}) with a comparison to the numerical solution of the eigenvalue problem. The left panel corresponds to the high-frequency regime, with p-modes, and the right panel shows the gravitationally unstable mode. The gray curves indicate how this mode emerges as the gravitational dilution factor $\eta$ is increased, i.e. when going from the Cowling approximation to taking gravity fully into account.
The present simplistic example is only meant to illustrate how one can derive dispersion relations from the formalism presented in this paper, but the latter is adapted for non-uniformly magnetized (along the plane) polytropic fluids.}
\label{fig:Example}
\end{figure*}

Expression \eqref{omega2GI_thinLimit} looks similar to the homogeneous Jeans criterion $\omega^2 = c^2 k_0^2 - \omega_c^2$ (taking the average density $\bar{\rho}$ equal to $\rho_c$), but it is in fact extremely different. Indeed, compared to the homogeneous case, the largest scales are stabilized, so that there exists a non-vanishing wavenumber of maximal growth rate.
Staying up to second order in $x_b$, it appears that instability ($\omega^2<0$) occurs only at wavenumbers smaller than the critical wavenumber
\begin{equation}
k_\text{crit}= \eta \ \! \bar{x}_b  \ \! k_\textsc{J},
\label{ks_thin_limit}
\end{equation}
where $\bar{x}_b \equiv x_b k_\textsc{J}$ is the dimensionless half-thickness of the slab, and the growth will be maximal for the wavenumber
\begin{equation}
k_\text{max}=\frac{k_\text{crit}}{2},
\end{equation}
at a rate given by
\begin{equation}
\omega_\text{max}^2=- \frac{\eta^2 \bar{x}_b^2}{4} \ \omega_c^2.
\end{equation}
The fact that in \eqref{omega2GI_thinLimit} the term $- \eta \omega_c^2 x_b k_0$ responsible for destabilization is proportional to the gravitational dilution factor $\eta$ makes it fully explicit that Jeans' instability is directly stemming from the last term in the force operator $\vec{F}$ given by \eqref{ForceOperator}.
Expression \eqref{omega2GI_thinLimit} also introduces an effective sound speed, but the departure from the usual $c^2$ is small since the term containing $x_b^2$ is of second order in this thin limit.
We plot these results in figure~\ref{fig:Example}, where the new feature compared to other studies is that we have the explicit dependency with $\eta$ which switches continuously from the Cowling case to the full case.

Finally, to compare expression \eqref{omega2GI_thinLimit} to the literature, let us detail how \cite{KimEtAl12} obtain their approximate analytical results. To begin with, they integrate along the stratification the mass conservation equation \eqref{AsAFunctionOfXi_rho1} and the momentum equation \eqref{VectorEigenvalueProblem} perpendicular to the stratification, to express the parameter $\omega^2$ from the left-hand side of the momentum equation \eqref{VectorEigenvalueProblem} as
\begin{equation}
\omega^2 = c^2_\text{eff} k_0^2 + g_\text{eff} \ \! k_0,
\label{DispersionRelationKimEtAl12}
\end{equation}
with an effective speed of sound and effective gravitational acceleration, respectively given by\footnote{Setting $\eta=1$ here to compare with their work where they do not introduce this dilution factor.}
\begin{equation}
\renewcommand\arraystretch{1.8} 
\begin{array}{l}
\displaystyle
c^2_\text{eff} \equiv \frac{\Sigma_0}{\Sigma_1} \ \! \langle \frac{p_1}{\rho_0} \rangle, \\
\displaystyle
g_\text{eff} \equiv \frac{\Sigma_0}{\Sigma_1} \ \! \langle i \ \! g_{1y} \rangle,
\end{array}
\label{DispersionRelationKimEtAl12_Coefficients}
\end{equation}
where $\Sigma_0 \equiv \int_{-x_b}^{x_b} \rho_0 \ \! \dd x$ is the equilibrium column density, $\Sigma_1 \equiv \int_{-x_b}^{x_b} \rho_1 \ \! \dd x$ is its perturbed version, and where the angles indicate density-weighted averages along the stratification, i.e. for a quantity $Q$ we have $\langle Q  \rangle \equiv \int_{-x_b}^{x_b} \rho_0 Q \ \! \dd x / \Sigma_0$. Of course, the simplicity of the dispersion relation \eqref{DispersionRelationKimEtAl12} is only apparent, since the perturbed quantities $\rho_1, p_1$ and $g_1$ entering the coefficients \eqref{DispersionRelationKimEtAl12_Coefficients} are unknown functions. However, by solving numerically the eigenvalue problem, \cite{KimEtAl12} notice that, in the cases they consider and as far as the fundamental mode is concerned, these functions are well approximated by simple functions. Notably in the isothermal case we have $\rho_1 \sim \rho_0$, $p_1 \sim c^2 \rho_0$, such that $c^2_\text{eff} \sim c^2$. As for $g_1$, they consider the integral form of Poisson equation with $\rho_1 \sim \rho_0$. Doing so they obtain results of the form $g_\text{eff} = - 2 \pi G \Sigma_0 k_0 \mathcal{F}$, where $\mathcal{F}$ is known in the literature as the gravity reduction factor. For the present case to which we want to compare our results (thin limit with rigid boundary conditions), $\mathcal{F} \rightarrow 1$. All in all, they obtain
\begin{equation}
\omega^2_\text{GI} = c^2 k_0^2 - \omega_c^2 x_b \ \! k_0,
\label{DispersionRelationKimEtAl12_Approx}
\end{equation}
where we have used $\Sigma_0 \sim 2 \rho_c x_b$ since the slab is thin. The only difference between \eqref{DispersionRelationKimEtAl12_Approx} and \eqref{omega2GI_thinLimit} (with $\eta=1$ as it should) is that in the latter a correction due to the thickness appears in the effective sound speed, but since it is only of second order in $x_b$, both expressions are consistent in the thin limit.

\section{Conclusion and prospects}
\label{sec:Conclusion}

Our work aims at studying the intricate interplay between magnetic fields and gravity in self-gravitating media in a rigorous manner. We have reformulated into various compact, classical forms the MHD wave equation governing the waves and instabilities (notably Jeans' gravitational instability) in non-uniformly magnetized self-gravitating polytropic static plasma slabs. These reformulations constitute a necessary step towards their understanding by means of MHD spectral theory.
Indeed, these various forms have prepared the ground for the following studies.

On a theoretical viewpoint, firstly, our coupled Sturm-Liouville form generalizes most naturally the form derived by \cite{Goedbloed71,GKP} for the MHD wave equation in the Cowling approximation. Therefore, by analogy, the most natural next step would be to identify the nature (genuine or apparent) of the singularities entering our set of equations, following for example \cite{Ince56} who details the method for fourth-order equations. The next important step would be to extend the oscillation theorem derived by \cite{GoedbloedSakanaka74}, which itself extends the classical Sturm-Liouville oscillation theorem. Alternatively, one may as in \cite{GKP} go back to the original vector eigenvalue problem, with a linear dependence on $\omega^2$, and exploit the self-adjointness of the force operator $\rho^{-1} \vec{F}$.
Secondly, our reformulation as a simple first order matrix differential equation brings us the full analytical expressions for the displacement vector and for the perturbed gravitational field. In fact, we have indeed obtained them explicitly in the thin limit already, but one may extend our expressions for arbitrarily thick slabs.

On a more practical viewpoint, by extending our simple illustration of section~\ref{sec:Example}, it is now straightforward to derive explicitly the dispersion relation of a polytropic magnetized self-gravitating slab in the thin limit. Considering notably free boundary conditions, one may then quantify analytically results obtained numerically in the literature \citep[e.g.][]{NagaiEtAl98}, completing them with the exhaustive understanding provided by spectral theory, and without assuming isothermality as is often done. Moreover, studying the singularities of the wave equation also yields practical stability criteria \citep[e.g.][]{Veugelen86}, in particular Suydam's criterion \citep[e.g.][]{Goedbloed73}. Finally, our coupled harmonic oscillator form, notably in its Hamiltonian form, opens up new prospects for stability analyses of this system, inspired from quantum mechanical oscillators and dynamical systems theory.

\vspace{-0.5cm}

\section*{Acknowledgments}

JBD thanks Ludovic Margerin for fruitful discussions. We thank the reviewer for his constructive comments.
This research is supported by the Agence Nationale de la Recherche (project BxB: ANR-17-CE31-0022).
RK and JBD are supported by Internal funds KU Leuven, project C14/19/089 TRACESpace. RK further received funding from the European Research Council (ERC) under the European Union's Horizon 2020 research and innovation programme (grant agreement No. 833251 PROMINENT ERC-ADG 2018) and a joint FWO-NSFC grant G0E9619N. 

\section*{Data availability}

No new data were generated or analysed in support of this research.

\vspace{-0.5cm}

\bibliographystyle{mnras}

\bsp

\bibliography{MagJeans}

\label{lastpage}

\appendix

\section{Derivation of the coupled Sturm-Liouville form}
\label{section:Appendix_Derivation_CSL_Form}

In this appendix we show how one may obtain the coupled Sturm-Liouville form \eqref{CSL} from the system \eqref{2ndOrderSystem}, i.e. we explain where our change of variable \eqref{def:chi} comes from.

First of all, let us do the change of variable
\begin{equation}
\varphi \equiv \frac{\mathcal{G}}{u},
\label{def:varphi}
\end{equation}
where $u$ is an unknown function at this stage. The idea is to rewrite our equations with this new degree of freedom, and then choose a constraint on $u$ which makes the equations more convenient. Noticing that we may write
\begin{equation}
\mathcal{G}'' = \frac{1}{u} (u^2 \varphi')' + u'' \varphi,
\end{equation}
and
\begin{equation}
\xi' = \frac{1}{\omega_\star^{2} u} [(\omega_\star^{2} u \xi)' - (\omega_\star^{2} u)' \xi],
\end{equation}
Poisson's equation \eqref{2ndOrderSystem_Poisson} becomes
\begin{equation}
[u^2 \varphi' + \omega_\star^{2} u \xi]' + u[u''-k_0^2 (1 + \eta \tfrac{\omega_\star^{2}}{\omega^2}) u] \varphi +  [k_\star \ \! \omega_\star^{2} u -(\omega_\star^{2} u)'] \xi = 0.
\label{IntermediatePoisson}
\end{equation}
This suggests, in order to get closer to a Sturm-Liouville form, to choose $u$ such that the last term vanishes, i.e. such that
\begin{equation}
k_\star \ \! \omega_\star^{2} u -(\omega_\star^{2} u)' = 0,
\label{def:u}
\end{equation}
that is, in integrated form
\begin{equation}
u(x) = \omega_\star^{-2} e^{\int k_\star dx}.
\label{u_exp}
\end{equation}
Then, in order to express \eqref{IntermediatePoisson} in a compact form, let us define\footnote{Tildes in $\widetilde{\chi}$ and $\widetilde{P}_2$ indicate that these are intermediate variables: we are going to change them, namely in \eqref{Final_chi} and \eqref{Final_P2}, to get the final $\chi$ and $P_2$ functions.}
\begin{equation}
\widetilde{\chi} \equiv u^2 \varphi' + \omega_\star^{2} u \xi,
\label{def:psi}
\end{equation}
and
\begin{equation}
\widetilde{P}_2 \equiv \frac{1}{u[u''-k_0^2 (1 + \eta \tfrac{\omega_\star^{2}}{\omega^2}) u]},
\label{def:delta}
\end{equation}
such that \eqref{IntermediatePoisson} simply reads
\begin{equation}
\widetilde{P}_2 \widetilde{\chi}' + \varphi = 0.
\label{Poisson_with_delta_psi_varphi}
\end{equation}
Now, since $\widetilde{\chi}$ is basically the derivative of $\varphi$, it is natural to differentiate \eqref{Poisson_with_delta_psi_varphi} to get an equation on $\widetilde{\chi}$ and $\xi$ only. Doing so, and using the definitions \eqref{def:u} and \eqref{def:psi}, we get that (the derivative of) the linearized Poisson equation can be written in the Sturm-Liouville form
\begin{equation}
\left(\widetilde{P}_2 \widetilde{\chi}'\right)' + u^{-2} \ \widetilde{\chi} = \omega_\star^{2} u^{-1} \ \xi.
\label{SL1}
\end{equation}
Likewise, let us write the momentum equation \eqref{2ndOrderSystem_MomentumConservation} with the same variable $\varphi$: from the definition \eqref{def:varphi} we get
\begin{equation}
\frac{\eta}{4 \pi G} \omega_\star^{2} u \varphi' + \left. \left. \frac{\eta}{4 \pi G} \right[(\omega_\star^{2} u)' - k_\star \ \! \omega_\star^{2} u \right] \varphi - (P\xi')'-Q\xi=0.
\label{IntermediateMomentumConversation}
\end{equation}
Now watch the miracle: From our choice \eqref{def:u} for the constraint on the function $u$, the second term on the left hand side of \eqref{IntermediateMomentumConversation} vanishes, such that, using the variable $\widetilde{\chi}$ from \eqref{def:psi}, equation \eqref{2ndOrderSystem_MomentumConservation} becomes
\begin{equation}
\left(P \xi'\right)' + \left(Q + \frac{\eta}{4 \pi G} \omega_\star^{4}\right) \xi = \frac{\eta}{4 \pi G} \omega_\star^{2} u^{-1} \widetilde{\chi},
\label{SL2}
\end{equation}
i.e. this particular choice of $u$ enables us to write Poisson's equation and the momentum equation \textit{simultaneously} in Sturm-Liouville forms.
A last convenient step is to notice that by working with the variable
\begin{equation}
\chi \equiv \sqrt{\frac{\eta}{4 \pi G}} (\omega_\star^{2} u)^{-1} \widetilde{\chi},
\label{Final_chi}
\end{equation}
and putting
\begin{equation}
P_2 \equiv \omega_\star^{4} u^2 \widetilde{P}_2,
\label{Final_P2}
\end{equation}
together with the relation \eqref{def:u}, the function $u$ disappears from equations \eqref{SL1} and \eqref{SL2}. This variable $\chi$ corresponds to the variable introduced in \eqref{def:chi}, and equations \eqref{SL1} and \eqref{SL2} correspond to the coupled Sturm-Liouville form \eqref{CSL}.

\section{Expliciting the solutions}
\label{appendix:ExplicitingSolutions}

Having reformulated our problem in classical forms enables us to express the solutions by means of standard expansions. For example, the solution of the $4 \times 4$ matrix form \eqref{CHO_4times4_MatrixForm} can be expressed as the Peano-Baker series
\begin{equation}
\renewcommand\arraystretch{2.} 
\begin{array}{l}
\displaystyle
\vec{V}(x) = \left[1 \! \! 1 + \int_{0}^x \! \! dx_1 \ \! \mathsf{M}_\textsc{v}(x_1) \right. \\
\displaystyle
\hspace{1.cm} \left. + \! \! \int_{0}^x \! \! dx_1 \int_{0}^{x_1} \! \! dx_2 \ \! \mathsf{M}_\textsc{v}(x_1) \ \! \mathsf{M}_\textsc{v}(x_2) + \dots\right] \vec{V}\left(0\right)
\end{array}
\label{PeanoBaker}
\end{equation}
and manipulating this form is greatly eased by the fact that $\mathsf{M}_\textsc{v}$ is a block matrix. Per se this is a formal solution in the sense that it is not guaranteed a priori that it converges or that the terms are ordered. However considering \eqref{PeanoBaker} truncated at first and second order already gives very good approximations to numerical solutions, but we will not explore this further here. Another idea is to solve perturbatively our problem using as perturbation parameter $\eta$, in the line of \cite{BenderOrszag78} for example, since the unperturbed problem is the well understood Cowling case. However, let us consider a third interesting expansion, namely expanding in terms of the thickness of the slab, as follows.

The equation \eqref{CHO_4times4_MatrixForm} is hard to solve because the matrix $\mathsf{M}_\textsc{v}$ is position dependent. However, if the slab thickness is smaller than the typical scale of variation of $\mathsf{M}_\textsc{v}$, the slab will be thin enough for this matrix to be roughly constant. The solution is then given by \eqref{V_ThinLimit_MatrixForm}. This observation calls for studying the solutions perturbatively using the thickness as perturbation parameter. To find a relevant dimensionless parameter for this,
we note that from the homogeneous case \citep[e.g.][]{Thompson06} it is well known that a key length scale in gravitational instability is given by the critical Jeans wavenumber
\begin{equation}
k_\textsc{J} \equiv \sqrt{\frac{4 \pi G \rho_0(x=0)}{c_a^2(x=0)}}.
\end{equation}
This length marks the balance between pressure gradients and the gravitational force. In principle $k_J$ is a position-dependent quantity because we do not assume a homogeneous equilibrium density profile $\rho_0$, but since we are about to expand our solution starting from a thin slab, it is most natural to use the central value (at $x=0$) of the Jeans wavenumber.
Let us define dimensionless positions as $\bar{x} \equiv k_\textsc{J} x$, and in particular the dimensionless position of the boundary
\begin{equation}
\bar{x}_b \equiv k_\textsc{J} x_b,
\end{equation}
which we use as perturbation parameter. Indeed, let us rewrite equation \eqref{CHO_4times4_MatrixForm} in dimensionless form
\begin{equation}
\frac{\dd V}{\dd \bar{x}} = \mathsf{M}_\textsc{v}(\bar{x}) V,
\label{CHO_4times4_MatrixForm_dimensionless}
\end{equation}
where we expand $\mathsf{M}_\textsc{v}$ as
\begin{equation}
\mathsf{M}_\textsc{v}(\bar{x}) = M_0(\bar{x}) + \bar{x}_b M_1(\bar{x}) + \mathcal{O} \left(\bar{x}_b^2\right),
\label{Mperturbatif}
\end{equation}
and, following for instance \cite{Holmes13}, let us look for solutions of the form
\begin{equation}
V(\bar{x}) = V_0(\bar{x}) + \bar{x}_b V_1(\bar{x}) + \mathcal{O} \left(\bar{x}_b^2\right).
\label{Vperturbatif}
\end{equation}
Then equation \eqref{CHO_4times4_MatrixForm_dimensionless} reads at zeroth order in $\bar{x}_b$
\begin{equation}
\frac{\dd V_0}{\dd \bar{x}} = M_0 V_0,
\label{VprimeEqualAV_0thOrder}
\end{equation}
and at first order
\begin{equation}
\frac{\dd V_1}{\dd \bar{x}} = M_0 V_1 + M_1 V_0.
\label{VprimeEqualAV_1stOrder}
\end{equation}
Now by construction of the expansion \eqref{Mperturbatif}, $M_0(\bar{x})$ equals $\mathsf{M}_\textsc{v}(\bar{x})$ for $\bar{x}_b=0$. But in this case the slab is infinitely thin, therefore in fact $M_0(\bar{x})=\mathsf{M}_\textsc{v}(0)$, i.e. $M_0$ is a constant. For this reason the two above equations may be solved explicitly, as follows.

The solution of \eqref{VprimeEqualAV_0thOrder} is the matrix exponential
\begin{equation}
V_0(\bar{x}) = e^{\bar{x} \mathsf{M}_\textsc{v}(0)} V_0(0)
\label{V0_ExpAtmo}
\end{equation}
where  $V_0(0)$ is the value of the vector $V_0$ at the center, not to be confused with $V(0)$, both being linked by 
\begin{equation}
V(0)=V_0(0) + \bar{x}_b V_1(0),
\label{InitialCondition}
\end{equation}
which is \eqref{Vperturbatif} to first order in $\bar{x}_b$.
Equation \eqref{VprimeEqualAV_1stOrder} constitutes an inhomogeneous problem, i.e.\ with a source term. Namely, consider an equation of the form
\begin{equation}
\frac{dU}{dx} = B U + S(x)
\label{VprimeEqualAV_Inhomogeneous}
\end{equation}
with a given constant matrix $B$, a given $x$-dependent source term $S(x)$, and a given initial condition $U(0)$. Its solution is given by \citep[e.g.][]{Tracy16}
\begin{equation}
U(x) = e^{x B} U(0) + e^{x B} \int_0^x e^{-s B} S(s) ds.
\label{VprimeEqualAV_Inhomogeneous_Solution}
\end{equation}
Since equation \eqref{VprimeEqualAV_1stOrder} corresponds to \eqref{VprimeEqualAV_Inhomogeneous} with $U \equiv V_1$, $B \equiv \mathsf{M}_\textsc{v}(0)$ and $S(\bar{x}) \equiv M_1 V_0 = M_1 e^{\bar{x} M(0)} V_0(0)$ using \eqref{V0_ExpAtmo}, we obtain
\begin{equation}
V_1(\bar{x}) = e^{\bar{x} \mathsf{M}_\textsc{v}(0)} V_1(0) + e^{\bar{x} \mathsf{M}_\textsc{v}(0)} \int_0^{\bar{x}} e^{-s \mathsf{M}_\textsc{v}(0)} M_1 e^{s \mathsf{M}_\textsc{v}(0)} V_0(0) ds.
\label{V1_ExpAtmo}
\end{equation}
Here too, beware of the initial conditions: $V_1(0)$ is the value of the vector $V_1$ at the center, not to be confused with $V(0)$, both being linked by relation \eqref{InitialCondition}. All that is left to do now is to plug in \eqref{Vperturbatif} the expressions of $V_0(\bar{x})$ and $V_1(\bar{x})$ just deduced. Doing so, we shall use the initial condition \eqref{InitialCondition}, in particular to replace the $V_0(0)$ vector in \eqref{V1_ExpAtmo} by $V(0)$, since we are working up to order one in $\bar{x}_b$. All in all, for a `thin slab', defined as $\bar{x}_b \ll 1$, we have
\begin{equation}
V(\bar{x}) = e^{\bar{x} \mathsf{M}_\textsc{v}(0)} \left[1 \! \! 1 + \bar{x}_b \int_0^{\bar{x}} Z_1(s) ds\right] V(0)
\label{ExpAtmo_Vtot}
\end{equation}
where
\begin{equation}
Z_1(s) \equiv e^{-s \mathsf{M}_\textsc{v}(0)} M_1 e^{s \mathsf{M}_\textsc{v}(0)}.
\label{ExpAtmo_Z1}
\end{equation}
This procedure may be pushed to arbitrary orders. At order two, the same steps yield
\begin{equation}
\renewcommand\arraystretch{2.3} 
\begin{array}{l}
V(\bar{x}) = e^{\bar{x} \mathsf{M}_\textsc{v}(0)} \left[1 \! \! 1 + \bar{x}_b \int_0^{\bar{x}} ds_1 Z_1(s_1)\right. \\
\left.+ \bar{x}_b^2 \left[\int_0^{\bar{x}} ds_1 \int_0^{s_1} ds_2 Z_1(s_1) Z_1(s_2) + \int_0^{\bar{x}} ds_1 Z_2(s_1)\right] \right] V(0),
\end{array}
\end{equation}
and in fact all orders are a sum of products of $Z_n(s) \equiv e^{-s \mathsf{M}_\textsc{v}(0)} M_n e^{s \mathsf{M}_\textsc{v}(0)}$. Hence we may get the solution beyond the regime $\bar{x}_b \ll 1$. Note that this infinite expansion is very different from the formal solution \eqref{PeanoBaker} because now the terms are ordered, with respect to the parameter $\bar{x}_b$, while in \eqref{PeanoBaker} we do not control a priori the amount of information lost when stopping the expansion at a finite order.

\section{Getting back to the initial variables $\xi$ and $\mathcal{G}$ once $\chi_1$ and $\chi_2$ are obtained}
\label{appendix:BackToInitialVariables}

In our derivation, we introduce the variables $\chi_1$ and $\chi_2$ which make the calculations far more convenient. But once the solution is found in terms of $\vec{V}=(\chi_1',\chi_2',\chi_1,\chi_2)^{\textsc{T}}$, as for example in the thin limit \eqref{V_ThinLimit_MatrixForm_Version1} or as detailed in appendix \ref{appendix:ExplicitingSolutions}, one needs to get back to the initial variables $\xi$ and $\mathcal{G}$, of physical interest. Obtaining their derivatives $\xi'$ and $\mathcal{G}'$ is also useful to deduce $g_x$ through \eqref{FieldLineProjection_of_irrotationalConstraint}, and $\xi_\parallel$ and $\xi_\perp$ through \eqref{XiPerp_XiParallel}.
For $\xi$ and $\xi'$ this is straightforward, as the definition \eqref{CHO_def_chi1_chi2} of $\chi_1$ gives directly
\begin{equation}
\begin{array}{ll}
\xi = \frac{\chi_1}{\sqrt{|P_1|}},\\
\xi' = \frac{1}{\sqrt{|P_1|}} \left(\chi_1'-\frac{1}{2}\frac{P_1'}{P_1} \chi_1\right).
\end{array}
\end{equation}
We now detail the less obvious steps, regarding $\mathcal{G}$.

\underline{Obtaining $\mathcal{G}$ once $\chi$ and $\chi'$ are known}
From the definition \eqref{def:chi} we have an expression for $\chi$ in terms of $\xi$, $\mathcal{G}$ and $\mathcal{G}'$. Taking the derivative of this relation we get an expression for $\chi'$ which contains $\mathcal{G}''$. We eliminate $\mathcal{G}''$ using its expression from Poisson's equation \eqref{2ndOrderSystem_Poisson}. It is then easy to see that the combination $\chi'+k_\star \chi$ eliminates $\mathcal{G}'$, so that only a dependency on $\mathcal{G}$ remains. Inverting that relation we finally get
\begin{equation}
\mathcal{G}=\sqrt{\frac{4 \pi G}{\eta}} \frac{\chi'+k_\star \chi}{\frac{k_0^2}{\omega_\star^{2}}(1+\eta \frac{\omega_\star^{2}}{\omega^2})+\left(\frac{\omega_\star^{2'}}{\omega_\star^{4}}-\frac{k_\star}{\omega_\star^{2}}\right)'+k_\star \left(\frac{\omega_\star^{2'}}{\omega_\star^{4}}-\frac{k_\star}{\omega_\star^{2}}\right)},
\end{equation}
and from the definition \eqref{CSL_Coeffs_2} of $P_2$ we notice that this can in fact be written simply as
\begin{equation}
\mathcal{G}=-\sqrt{\frac{4 \pi G}{\eta}} \frac{P_2}{\omega_\star^{2}} (\chi'+k_\star \chi).
\label{G_inTermsOf_chi_and_chiprime}
\end{equation}

\underline{Obtaining $\mathcal{G}$ once $\chi_2$ and $\chi_2'$ are known}
Similarly, taking the derivative of the definition \eqref{CHO_def_chi1_chi2} of $\chi_2$, and using the expression of $\chi'$ in terms of $\xi$, $\mathcal{G}$ and $\mathcal{G}'$ deduced above, one can see that the combination $\chi_2'-(\tfrac{P_1'}{2 P_1}-k_\star) \chi_2$ eliminates $\mathcal{G}'$ and depends on $\mathcal{G}$ only. Solving for $\mathcal{G}$, and using the definition \eqref{CSL_Coeffs_2} of $P_2$, we thus get
\begin{equation}
\mathcal{G}= \sqrt{\frac{4 \pi G}{\eta}} \frac{\text{sign}(P_2) \sqrt{|P_2|}}{\omega_\star^{2}} \left[(\tfrac{P_1'}{2 P_1}-k_\star) \chi_2-\chi_2'\right].
\label{G_inTermsOf_chi2_and_chi2prime}
\end{equation}

\underline{Obtaining $\mathcal{G}'$ once $\chi_1$ and $\chi_2$ are known}
From the definition \eqref{def:chi} of $\chi$ we may express $\mathcal{G}'$ in terms of $\chi$, $\xi$ and $\mathcal{G}$. Together with the definition \eqref{CHO_def_chi1_chi2} of $\chi_1$ and $\chi_2$, this reads
\begin{equation}
\mathcal{G}'= \omega_\star^{2} \left[\sqrt{\frac{4 \pi G}{\eta}} \frac{\chi_2}{\sqrt{|P_2|}}-\frac{\chi_1}{\sqrt{|P_1|}}\right]+\left(k_\star-\frac{\omega_\star^{2'}}{\omega_\star^{2}}\right) \mathcal{G}.
\label{Gprime_InTermsOf_G}
\end{equation}
Injecting \eqref{G_inTermsOf_chi2_and_chi2prime} in the above expression, we get $\mathcal{G}'$ in terms of $\chi_1$, $\chi_2$ and $\chi_2'$.

\ \\
Finally, we gather the above expressions into the matrix form
\begin{equation}
\left(
\begin{array}{c}
\mathcal{G}' \\
\mathcal{G} \\
\xi'\\
\xi
\end{array}
\right)
=
\mathsf{M}_3
\left(
\begin{array}{c}
\chi_1' \\
\chi_2' \\
\chi_1\\
\chi_2
\end{array}
\right),
\end{equation}
where the matrix $\mathsf{M}_3$ is given in \eqref{def:matrix_M3}. The above relation allows us to obtain the final form \eqref{def:xi_components_FINALRESULT} for the displacement vector $\vec{\xi}$.

\end{document}